\documentclass[11pt]{article}
\usepackage{graphicx,rotating,hyperref,slashed,amsmath,xcolor,amssymb,amsfonts,colortbl,cite,soul,cancel,cleveref,framed,ulem} 
\usepackage{graphicx} 
\usepackage{dsfont}
\usepackage{hyperref}
\usepackage{cleveref}
\usepackage{enumerate}

\usepackage[export]{adjustbox}
\makeatletter   
\hypersetup{colorlinks,bookmarksopen,bookmarksnumbered,
linkcolor=blue,pdfstartview=FitH,urlcolor=rossos,citecolor=verde}
\allowdisplaybreaks	 
\numberwithin{equation}{section} 
  \usepackage[paperwidth=7.5in, paperheight=11in, margin=1in]{geometry}

\hypersetup{%
    ,urlcolor=black
    ,citecolor=black
    ,linkcolor=black 
    }
\usepackage{mciteplus}

\AtBeginDocument{
  \hypersetup{
    urlcolor=blue,
    citecolor=blue,
    linkcolor=blue,
  }%
}
\newcommand{\sigwfast}{\texttt{SIGWFast}}
\newcommand{\sigway}{\texttt{SIGWAY}}
\newcommand{\zc}{\ensuremath{\mathcal{Z}}}
\newcommand{\lc}{\ensuremath{\mathcal{L}}}

\newcommand{\GW}{\ensuremath{\mathrm{GW}}}

\usepackage{setspace} 

\def\be{\begin{equation}}
\def\ee{\end{equation}}

\def\bea{\begin{eqnarray}}
\def\eea{\end{eqnarray}}

\def\T{{\cal T}}

\def\mP{\mathcal{P}}

\def\0{{\boldsymbol 0}}

\begin{document}

\begin{titlepage}

\setcounter{page}{1} \baselineskip=15.5pt \thispagestyle{empty}
\bigskip\

\vspace{1cm}
\begin{center}
\baselineskip=1.5\baselineskip
{\fontsize{20}{32}\selectfont  \sffamily \bfseries {Bayesian reconstruction of primordial perturbations from induced gravitational waves}}
\end{center}

\vspace{-0.1cm}

\renewcommand{\thefootnote}{\fnsymbol{footnote}}
\begin{center} 
{\fontsize{13}{30}
 Aya Ghaleb$^{a,\!}$~\footnote{\texttt{aya-ghaleb@outlook.com}}, Ameek Malhotra$^{a,\!}$~\footnote{\texttt{ameek.malhotra@swansea.ac.uk}}, Gianmassimo Tasinato$^{a,b\!}$~\footnote{\texttt{g.tasinato2208@gmail.com}}, Ivonne Zavala$^{a,\!}$~\footnote{\texttt{e.i.zavalacarrasco@swansea.ac.uk}}
} 
\end{center}

\begin{center}
\vskip 6pt
\textsl{$^a$ Physics Department, Swansea University, SA2 8PP, UK}
\\
\textsl{$^{b}$ Dipartimento di Fisica e Astronomia, Universit\`a di Bologna,\\
 INFN, Sezione di Bologna,  viale B. Pichat 6/2, 40127 Bologna,   Italy}
\vskip 4pt
\end{center}

\vspace{1.cm}
\hrule \vspace{0.3cm}
\noindent
The formation of primordial black holes or other dark matter relics 
from amplified density fluctuations in the early universe may also generate scalar-induced gravitational waves (GW), carrying vital information about the primordial power spectrum and the early expansion history of our universe. We present a Bayesian approach aimed at reconstructing both the shape of the scalar power spectrum and the universe's equation of state from GW observations, using interpolating splines to flexibly capture features in the GW data. The optimal number of spline nodes is chosen via Bayesian evidence, aiming at balancing complexity of the model and the fidelity of the reconstruction.  
We test our method using both representative mock data and recent Pulsar Timing Array measurements, demonstrating that it can accurately reconstruct the curvature power spectrum as well as the underlying equation of state, if  different from radiation. 
\vskip 10pt
\hrule
 
\vspace{0.4cm}
\end{titlepage}


\renewcommand{\thefootnote}{\arabic{footnote}}
\setcounter{footnote}{0}

\section{Introduction}
\label{sec:intro}

The power spectrum of curvature perturbations,  $\mP_\zeta(k)$, plays a crucial role in understanding the early Universe and the formation of cosmic structure. While the cosmic microwave background (CMB) and large-scale structure of the universe provide constraints on its features at large cosmological scales (see e.g. \cite{Planck:2018vyg}), smaller scales remain relatively unconstrained. One promising avenue to probe this regime  is through scalar-induced gravitational waves (SIGWs), which arise as a second-order effect of primordial density perturbations: see
e.g.  
\cite{Matarrese:1992rp,Ananda:2006af,Baumann:2007zm,Saito:2009jt,Bugaev:2009zh,Assadullahi:2009jc,Alabidi:2012ex,Alabidi:2013lya,Espinosa:2018eve,Kohri:2018awv}  
for  original and key papers, and \cite{Domenech:2021ztg} for a comprehensive,
detailed review. Such induced gravitational waves carry valuable information about the small-scale features of $\mP_\zeta(k)$ and may be detected by future gravitational wave observatories, such as the Laser Interferometer Space Antenna (LISA) \cite{LISA:2024hlh} and the Einstein Telescope (ET) \cite{Abac:2025saz}. Enhanced scalar fluctuations, leading
to observable GW signals, can be produced
on models leading to primordial black holes (PBH), see e.g. \cite{Ozsoy:2023ryl} for a review.
In addition to encoding information about the primordial power spectrum, the gravitational wave (GW) spectrum is also sensitive to the expansion history of the Universe (see e.g. \cite{Domenech:2019quo,Hajkarim:2019nbx,Domenech:2020kqm}). 
This is good news since deviations from a purely radiation-dominated phase at very early times can leave distinct imprints on the shape and amplitude of the GW spectrum, which
can be inferred from data.
Such departures might arise from exotic physics or alternative cosmological scenarios, and they can be parametrized in terms of an effective equation of state parameter \cite{Domenech:2021ztg}. A detection of SIGWs, therefore, offers a unique opportunity not only to reconstruct $\mP_{\zeta}(k)$, but also to infer the background expansion dynamics. 

\smallskip

The formation of 
SIGW is a process occurring at second order in fluctuations, and its difficult to analytically disentangle properties of the primordial sources from GW data alone. In this work, we  develop
an automated and efficient  method to do so, based on Bayesian inference techniques.  Our reconstruction framework 
allows also 
to simultaneously determine the equation of state parameter alongside the power spectrum, aiming to uncover hints of non-standard early universe evolution.

In our  Bayesian approach to reconstruct the power spectrum,
we represent it in the form of an interpolating spline with variable number of nodes, positions and amplitudes, all to be determined from GW data. For a model with fixed number of nodes, the posterior distributions of the spline position and amplitude parameters are determined using nested sampling while the Bayesian evidence criterion is used to compare the relative probabilities of models with different number of nodes. 
To test our method, we generate mock $\Omega_{\GW}$ for power spectra representative of early universe scenarios leading to enhanced scalar perturbations and three scenarios of post-inflationary cosmological background evolution -- radiation domination, early matter domination transitioning to radiation domination and a general equation of state $w$ different from matter or radiation. For the latter, we also test the viability of our method in determining $w$, 
 in addition to the curvature power spectrum. The results that we present here can be reproduced using our code available on GitHub.\footnote{\url{https://github.com/Ameek94/SIGW_Inverse}.\label{ftn_git}} 

\smallskip

This paper is organised as follows.
Section \ref{sec:sigwkernels} reviews  theory aspects of SIGW.
In \cref{sec:method} we describe our method, starting from the calculation of $\Omega_{\GW}$  to generate our mock data and then discuss in detail the inverse problem of reconstructing $\mP_{\zeta}$ from $\Omega_{\GW}$. \Cref{sec:results} presents the results of our reconstruction  for  representative examples as well as using recent Pulsar Timing Array data. \Cref{sec:conclusions} presents our conclusions, followed
by technical appendixes.

\section{Induced GWs and  the Universe's expansion history} \label{sec:sigwkernels}

In this section  we collect and outline the key equations and conventions  relevant for the computation of scalar-induced gravitational waves (SIGWs) in different cosmological scenarios. We review three cases for the post-inflationary evolution:  the standard case of radiation domination (RD) $c_s^2=w=1/3$ (see the review \cite{Domenech:2021ztg}), 
a sudden transition from an early matter-dominated (eMD) phase to RD (see e.g. \cite{inomata_enhancement_2019}), and
an adiabatic perfect fluid domination with  general (constant) equation of state $c_s^2=w$ \cite{Domenech:2021ztg}.
These three cases will then be analysed
respectively in sections \ref{sec:RD}, \ref{sec_EMD}, and \ref{sec:gen_w}.

The physics of SIGWs relies on second-order effects in cosmological perturbation theory. Primordial GW are amplified once scalar curvature perturbations originating from quantum fluctuations during inflation act as sources for tensor modes (GW) upon their horizon re-entry. The detectability of the resulting gravitational wave background relies in part on the amplitude of these scalar perturbations: a sufficiently enhanced scalar power spectrum at small scales can lead to observable SIGWs in current or future gravitational wave observatories.

On large cosmological scales, such as those probed by the cosmic microwave background (CMB) with characteristic wavenumbers $k \sim 0.001 \mathrm{Mpc}^{-1}$, observations constrain the amplitude of scalar perturbations to be of order $\mathcal{P}_\zeta \sim 10^{-9}$ \cite{Planck:2018vyg}. This leads to an unobservable small SIGW signal on CMB scales. At much smaller scales ($k \gg 1 \mathrm{Mpc}^{-1}$), however, there are no direct observational bounds on $\mathcal{P}_\zeta$, allowing for the possibility of significant enhancement. In such scenarios, the induced gravitational wave spectrum can attain amplitudes within the sensitivity range of
a variety of GW detectors. This opens a new observational window into the small-scale structure of the primordial Universe, complementary to CMB and large-scale structure surveys, which are sensitive to larger scales.

The energy density spectrum of SIGWs arises from a convolution over two copies of the scalar power spectrum $\mathcal{P}_\zeta$ and a transfer function $\mathcal{T}$ (also referred to as the kernel), which encodes the evolution of the tensor modes in the cosmological background. The dominant contribution is given by a double integral over momenta, weighted by these two elements.

In the standard thermal history, inflation is followed by reheating, which initiates a radiation-dominated (RD) era. Scalar perturbations that exit the horizon during inflation re-enter during RD and source GWs at second order. However, SIGWs are not merely sensitive to the inflationary perturbations, they also offer a probe of the Universe’s post-inflationary thermal history, particularly the epoch between inflation and Big Bang Nucleosynthesis (BBN), which remains largely unconstrained by conventional observations. This period could include non-standard cosmological phases, such as early matter domination or kination, motivated by scenarios in high-energy physics and string theory\footnote{See \cite{Cicoli:2023opf} for a recent review on string cosmology, and \cite{Allahverdi:2020bys,Batell:2024dsi} for reviews  on non-standard post-inflationary histories.}. 

The energy density spectrum of SIGWs takes the general form \cite{Domenech:2021ztg}

\begin{equation} \label{eq:omega_def}
\Omega_{\mathrm{GW}}(\eta, k) = \frac{1}{12}\left(\frac{k}{\mathcal{H}(\eta)}\right)^2 \overline{\mathcal{P}_h(\eta, k)} ,
\end{equation}
where $\eta$ is the conformal time, $\mathcal{H}(\eta)=a^{\prime}(\eta)/a(\eta)$  the conformal Hubble parameter, and the overline denotes an average over oscillations. The induced tensor power spectrum can be written as:
\begin{equation} \label{eq:Ph_def}
\overline{\mathcal{P}_h(\eta, k)}=8 \int_0^{\infty} \mathrm{d}v \int_{|1-v|}^{1+v} \mathrm{d}u \left(\frac{4v^2 - (1 + v^2 - u^2)^2}{4uv}\right)^2 \overline{I^2(u, v, k, \eta)} , \mathcal{P}_{\zeta}(ku) \mathcal{P}_{\zeta}(kv) ,
\end{equation}
where $\mathcal{P}_{\zeta}$ is the curvature power spectrum, and
\begin{equation}\label{eq:momenta_def}
    v \equiv \frac{q}{k} \quad, \quad u \equiv \frac{|\mathbf{k}-\mathbf{q}|}{k} .
\end{equation}

The kernel $I$, defined by a convolution of the source with a Green's function, depends on the evolution of the scalar perturbations and the background cosmology.
\begin{equation}\label{eq:kernel_def}
I(u, v, k ,\eta)=\int_{\eta_i}^{\eta} \mathrm{d}\tilde{\eta}\,  G_k(\eta, \tilde{\eta})  f(u, v, k, \tilde{\eta}) \,.
\end{equation}

The exact form of this expression changes with the equation of state of the Universe, motivating us to now present the specific cases we analyse in what follows.


\begin{enumerate}[(a)]

    \item {\it Radiation domination}: This
    is the most well-studied
    case in the literature, see e.g. 
    \cite{Domenech:2021ztg}, and corresponds
    to an equation of state with $w=1/3$. The expression
    for the induced $\Omega_{\rm GW}$ results
   \begin{equation}\label{eq:Omega_rh_RD}
        \Omega_{\mathrm{GW}, \mathrm{RD}}=
        \int_0^{\infty} d v \int_{|1-v|}^{1+v} d u \mathcal{T}_{\rm RD}\left(u, v,c_s\right) \mathcal{P}_{\zeta}(k u) \mathcal{P}_{\zeta}(k v)\,,
    \end{equation}
where $c_s$ is the propagation speed of scalar perturbations and the function     $\T_{\rm RD}(u,v,c_s)$ 
    is given by
    \begin{eqnarray}
            \mathcal{T}_{\rm RD}& =&
    \frac{y^2}{3\,c_s^4}
    \left[
    \frac{4v^2 - \bigl(1 - u^2 + v^2\bigr)^2}{4\,u^2\,v^2}
    \right]^2 
    \nonumber
    \\
    &&
    \times \left(
    \frac{\pi}{2}\,4\,y^2\,\Theta\!\bigl[c_s(u+v)-1\bigr]
    +\Bigl(1 - \tfrac12\,y\ln\frac{1+y}{1-y}\Bigr)^2
    \right)\,, 
        \label{eq:TRD}
     \end{eqnarray}
where $y$ is defined as:
    \begin{equation}\label{y_def}
        y=1-\frac{1-c_s^2(u-v)^2}{2 c_s^2 u v}\,.
    \end{equation}
In our work we compute the GW density  $\Omega_{\mathrm{GW}}$ for this case
 using the numerical code \sigway \cite{LISACosmologyWorkingGroup:2025vdz}. 

     \item {\it Early matter domination}:
     We consider a scenario in which a period of early matter domination (eMD) precedes the onset of radiation domination (RD), with a sudden transition between the two phases—referred to here as reheating. A distinctive feature of this setup is the sharp enhancement of the scalar-induced gravitational wave (SIGW) signal immediately after the transition~\cite{inomata_enhancement_2019}. This enhancement arises because, during eMD, the scalar source term responsible for GW production remains constant and does not efficiently generate GWs. However, the onset of  radiation domination  
     leads to efficient GW production. This characteristic imprint in the GW spectrum may fall within the sensitivity reach of upcoming observatories such as LISA, providing a unique observational window into the pre-BBN universe and the reheating dynamics.

It is important to note that this enhancement is sensitive to the details of the transition. The assumption of a {\it sudden transition} corresponds to a change in the background evolution occurring on a timescale much shorter than the Hubble time at reheating. In realistic scenarios, however, the transition is expected to be gradual. During such an intermediate period, the evolution of perturbations is continuous, and their behaviour cannot be fully captured by either the pure eMD or RD approximations. Several studies have explored this regime using smoothed background interpolations~\cite{Inomata_gradual_2019,pearce2024InterpolatingFastandSlow,pearce2025EarlyMatterDominatedEpochs,Kumar:2024hsi}, showing that the resulting SIGW spectrum is modified in both amplitude and shape compared to the sudden-transition limit.
    
    As discussed in \cite{inomata_enhancement_2019}, assuming a sudden transition, however, remains a reasonable approximation in two key regimes: for modes with $k \ll k_{\mathrm{rh}}$, which remain superhorizon at reheating and thus have constant curvature perturbations unaffected by the nature of the transition; and for modes with $k \gg k_{\mathrm{rh}}$, which are already subhorizon and oscillating well before reheating, making their evolution similarly insensitive to how reheating proceeds. The approximation begins to fail, however, for modes with $k \sim k_{\mathrm{rh}}$, which enter the horizon around the time of reheating and are therefore more sensitive to whether the transition is sharp or gradual. That said, if the peak of the primordial power spectrum corresponds to modes that enter the horizon well before reheating, then the dominant contribution to the induced gravitational wave signal comes from earlier times. In this case, reheating does not introduce a significant new source of GWs, and only affects a narrow range of modes near $k_{\mathrm{rh}}$, which do not dominate the overall spectrum. For instance, in models with a sharp infrared (IR) cutoff at some scale $k_{\text {IR }}$, where $k_{\text {IR }} \gg k_{\text {rh }}$, all relevant modes contributing to GW production enter the horizon before reheating. As long as the evolution of induced GWs is also correctly connected across the transition, then we can adopt this sudden transition scenario. This approach is demonstrated by numerical results such as \cite{inomata_enhancement_2019} and \cite{LISACosmologyWorkingGroup:2025vdz}.

    The  kernel entering \cref{eq:Ph_def} can be decomposed into
    \be
    \overline{I^2\left(u, v, x, x_{\mathrm{rh}}\right)} \simeq \overline{I_{\mathrm{eMD}}^2\left(u, v, x, x_{\mathrm{rh}}\right)}+\overline{I_{\mathrm{RD}}^2\left(u, v, x, x_{\mathrm{rh}}\right)}\,,
    \ee
    where as before, a subscript `rh' denotes the time of reheating and  $x\equiv k \eta$. In sudden reheating scenarios, we focus on gravitational waves induced during the radiation-dominated era by scalar perturbations that previously experienced an early matter-dominated phase on subhorizon scales. While GWs can also be generated during the eMD era \cite{Alabidi_2013, Assadullahi_2009}, their amplitude is typically much smaller than those produced during the RD era, where conditions are more favourable for GW amplification. This is particularly the case in sudden reheating scenarios \cite{inomata_enhancement_2019}. Thus, we consider primarily the RD contribution, and modes with $k>k_{\mathrm{rh}}$, which experience  a non-trivial evolution before the universe transitioned to radiation domination. The kernel is approximated in terms of two dominant contributions given in terms of the more convenient  variables, $s,t$ defined as 
    \be
    s\equiv u-v \,, \qquad t \equiv u+v -1\,.
    \ee  

    The first contribution is associated with modes deep inside the horizon, where the integral is dominated by the large scale  $t$ region (LS). In this limit, an analytical approximation is obtained by fixing $s=0$ and integrating over $t$, leading to
    \be
    \left.\overline{I_{\mathrm{RD,LS}}^2}\right|_{s=0} \simeq \frac{9 t^4 x_{\mathrm{rh}}^8\left(4 \mathrm{Ci}\left(\frac{x_{\mathrm{rh}}}{2}\right)^2+\left(\pi-2 \operatorname{Si}\left(\frac{x_{\mathrm{rh}}}{2}\right)\right)^2\right)}{81920000} \,.
    \ee
     This approximation is valid for large-scale modes with $k \ll k_{\mathrm{cut} }$, where $k_{\mathrm{cut}}$  is the scale corresponding to the onset of eMD, or the scale where density perturbations become non-linear \cite{inomata_enhancement_2019,Inomata_gradual_2019}. 
    
    The second contribution accounts for the resonance-like peak in the spectrum (res). In this case, the integration is performed over $s$, while fixing $t=\sqrt{3}-1$ throughout, except in the argument of the cosine integral function ${\rm Ci}$, where a logarithmic singularity leads to an enhancement:
    \be
    \overline{I_{\mathrm{RD,res}}^2} \approx Y \frac{9\left(-5+s^2+2 t+t^2\right)^4 x_{\mathrm{rh}}^8}{81920000(1-s+t)^2(1+s+t)^2} \mathrm{Ci}(|y|)^2\,,
    \ee
    where $y \equiv(t-\sqrt{3}+1) x_{\mathrm{rh}} /(2 \sqrt{3})$ and $Y$ is a numerical factor that absorbs the uncertainties in the integration limits. The total induced GW spectrum after reheating can be approximated as the sum of these two contributions:
    \be
    \Omega_{\mathrm{GW,rh}}^{\mathrm{eMDRD}} \simeq \Omega_{\mathrm{GW,rh}}^{(\mathrm{LS})}+\Omega_{\mathrm{GW,rh}}^{(\mathrm{res})} \,.
    \ee

In our work, by means of these analytical forms of the kernel,   we compute the GW density  $\Omega_{\mathrm{GW}}$ using the numerical code \sigway \cite{LISACosmologyWorkingGroup:2025vdz}.

    \item {\it General $w$}: 
    During a cosmological epoch characterised by a constant equation of state parameter $w$ and a constant scalar sound speed $c_s$, the transition to radiation domination introduces a new characteristic scale. This scale is defined by the comoving wavenumber $k_{\mathrm{rh}}$, which corresponds to the  mode that re-entered the horizon at the onset of radiation domination.

    The gravitational wave (GW) spectrum under such conditions has been studied  in
    \cite{Domenech:2019quo,Hajkarim:2019nbx,Domenech:2020kqm} 
    (see the review
    \cite{Domenech:2021ztg} for a detailed discussion). We are
    interested in particular to the regime where all relevant modes re-enter the horizon well before reheating, i.e., $k \gg k_{\mathrm{rh}}$. 
     The kernel averaged over oscillations is given by:
    \bea\label{eq:kernel_general}
    \begin{aligned} 
    \overline{I^2( u, v, x)}= & x^{-2(b+1)} 4^{2 b} \Gamma^4[b+3 / 2]\left(\frac{2 b+3}{b+2}\right)^2 \frac{\left|1-y^2\right|^b}{2 c_s^4 u^2 v^2} \\
    & \times\left\{\left(\mathrm{P}_b^{-b}(y)+\frac{b+2}{b+1} \mathrm{P}_{b+2}^{-b}(y)\right)^2 \Theta\left[c_s(u+v)-1\right]\right. \\
    & +\frac{4}{\pi^2}\left(\mathrm{Q}_b^{-b}(y)+\frac{b+2}{b+1} \mathrm{Q}_{b+2}^{-b}(y)\right)^2 \Theta\left[c_s(u+v)-1\right] \\
    & \left.+\frac{4}{\pi^2}\left(\mathcal{Q}_b^{-b}(-y)+2 \frac{b+2}{b+1} \mathcal{Q}_{b+2}^{-b}(-y)\right)^2 \Theta\left[1-c_s(u+v)\right]\right\}\,.
    \end{aligned}
    \eea
    Here, $\Theta[x]$ denotes the Heaviside step function. The functions $\mathrm{P}_b^{-b}(y)$ and $\mathrm{Q}_b^{-b}(y)$ are Ferrers functions of the first and second kind, valid for $|y|<1$, while $Q_b^{-b}(y)$ is the associated Legendre function of the second kind, used for $|y|>1$, 
    where  $y$ is defined as in \cref{y_def}.      
    This parameter is geometrically related to the shape of the momentum triangle formed by the interacting modes. Finally $b$ is related to the equation of state by: 
    \be\label{eq:bdef}
    b=\frac{1-3 w}{1+3 w} \,.
    \ee
 The presence of the Heaviside functions in eq.~\eqref{eq:kernel_general} indicates momentum conservation.

    For $k \gg k_{\mathrm{rh}}$, induced GWs are effectively free to propagate close to the reheating epoch. 
    Therefore, it is appropriate to evaluate the GW energy density spectrum at the moment of reheating, $\eta=\eta_{\mathrm{rh}}$, as:
    \begin{equation} \label{eq:Omega_matched}
        \left.\Omega_{\mathrm{GW}}\left(\eta \gg \eta_{\mathrm{rh}}, k \gg k_{\mathrm{rh}}\right)=\frac{1}{12}\left(\frac{k}{\mathcal{H}(\eta)}\right)^2 \overline{\mathcal{P}_h(\eta, k)} \right\rvert_{\eta=\eta_{\mathrm{rh}}}
    \end{equation}
    Remembering that $x=k\eta$, setting 
    \begin{equation}\label{eq:reheating_k}
        k_{\mathrm{rh}}=\mathcal{H}_{\mathrm{rh}}=\frac{1+b}{\eta_{\mathrm{rh}}}
    \end{equation}
    yields a scaling factor $\left(k / k_{\mathrm{rh}}\right)^{-2 b}$, accounting for the different redshifting behavior of GWs before radiation domination and ensuring continuity of the solution at $\eta=\eta_{\mathrm{rh}}$.
    
    Finally, the general expression for the GW spectrum at reheating, valid for any shape of the primordial spectrum as long as $k \gg k_{\mathrm{rh}}$, is given by
    \begin{equation}\label{eq:Omega_rh_w}
        \Omega_{\mathrm{GW}, \mathrm{rh}}=\left(\frac{k}{k_{\mathrm{rh}}}\right)^{-2 b} \int_0^{\infty} d v \int_{|1-v|}^{1+v} d u \mathcal{T}\left(u, v, b, c_s\right) \mathcal{P}_{\zeta}(k u) \mathcal{P}_{\zeta}(k v)
    \end{equation}
    where  the transfer function is defined as
    \begin{equation}
    \label{eq:Tgenw}
        \begin{aligned}
        \mathcal{T}\left(u, v, w\right)= & \mathcal{N}\left(b, c_s\right)\left(\frac{4 v^2-\left(1-u^2+v^2\right)^2}{4 u^2 v^2}\right)^2\left|1-y^2\right|^b \\
        & \times\left\{\left(\mathrm{P}_b^{-b}(y)+\frac{b+2}{b+1} \mathrm{P}_{b+2}^{-b}(y)\right)^2 \Theta\left[c_s(u+v)-1\right]\right. \\
        & +\frac{4}{\pi^2}\left(\mathrm{Q}_b^{-b}(y)+\frac{b+2}{b+1} \mathrm{Q}_{b+2}^{-b}(y)\right)^2 \Theta\left[c_s(u+v)-1\right] \\
        & \left.+\frac{4}{\pi^2}\left(\mathcal{Q}_b^{-b}(-y)+2 \frac{b+2}{b+1} \mathcal{Q}_{b+2}^{-b}(-y)\right)^2 \Theta\left[1-c_s(u+v)\right]\right\},
        \end{aligned}
    \end{equation}
    
    with the numerical coefficient given by
    \begin{equation} \label{eq:norm_coeff}
        \mathcal{N}\left(b, c_s\right) \equiv \frac{4^{2 b}}{3 c_s^4} \Gamma^4\left[b+\frac{3}{2}\right]\left(\frac{b+2}{2 b+3}\right)^2(1+b)^{-2(1+b)}\,.
    \end{equation}
   
    Following the early constant-$w$ epoch, the gravitational potential decays, and the GW spectrum will eventually become constant. So far we denoted the spectrum at this time by a subscript ‘rh’ and now relate it to the present-day amplitude, denoted by a subscript ‘0’, through standard cosmological evolution:
   \begin{equation} \label{eq:Omega_today}
        \Omega_{\mathrm{GW},0}=0.39\left(\frac{g_c}{106.75}\right)^{-1 / 3} \Omega_{\mathrm{r}, 0} \Omega_{\mathrm{GW}, \mathrm{rh}}
    \end{equation}
    where $\Omega_{\mathrm{r}, 0}$ is the current radiation energy density parameter, and $g_c$ is the effective number of relativistic degrees of freedom at $\eta=\eta_{\mathrm{rh}}$.

    In our analysis, we use a modified version of the public code \texttt{SIGWFast}~\cite{Witkowski:2022mtg} to compute the spectrum $\Omega_{\mathrm{GW}}(f)$\footnote{Where we  convert the wavenumber to  frequency using ${k}/{\text{Mpc}^{-1}} \simeq 6.5\times 10^{14} f/\mathrm{Hz}.$}\label{footfk},     
 given an input curvature power spectrum, a constant equation of state $w$, and a transition (or reheating) scale $f_{\text {rh }}$. The  \texttt{SIGWFast}  code follows the same conventions which we adopt throughout this work \footnote{Up to a minor redefinition of the variables $u, v$ in terms of $s, d$.}.
   
\end{enumerate}

\section{The reconstruction method}
\label{sec:method}

As reviewed in the previous section, the amplification
of primordial GWs by scalar non-linearities 
 is a complex process. The resulting
GW density $\Omega_{\rm GW}$ is typically expressed in terms
of convolution integrals depending on the (square of)
the curvature perturbation spectrum, as well as on
transfer functions of fluctuations through distinct cosmological
eras. As a consequence, it is not easy to analytically
disentangle the features of the scalar sources or
the cosmological expansion history from the resulting
frequency dependence of the GW density. In this section,
we introduce an efficient, automated method to do so, based
on Bayesian inference. The corresponding codes can be found
at the link in footnote \ref{ftn_git}. 

\subsection{Spline Parametrization of the Power Spectrum}
\label{sec_splmeth}

To reconstruct the primordial curvature spectrum $\mP_\zeta(k)$
starting from GW data, we model $\log_{10} \mP_\zeta(k)$ using splines, i.e.~a piecewise linear interpolation defined by a set of nodes, $\log_{10}\{(k_i, A_i)\}_{i=1}^N$, where $k_i$ denotes the wavenumber positions of the nodes and $\log_{10} A_i = \log_{10} \mP_\zeta(k_i)$ are the corresponding amplitudes. 
The linear spline method  provides a simple and flexible representation of the power spectrum,  capturing intricate features that may hide in the data. This approach is particularly advantageous compared to fixed functional forms (e.g.~power-law or broken power-law models), as it does not impose a specific shape on the power spectrum a priori. 
We now develop arguments based on Bayasian inference to select
the number and position of nodes.

\smallskip

The placement of the nodes is not predefined, but it is  treated as free parameter to be inferred from the data, ensuring that the model remains adaptable. Only the first and last node are held fixed, based on the range over which the curvature power spectrum is interpolated. This range is typically chosen to be slightly wider than the range over which the $\Omega_{\rm GW}$ is observed.\footnote{In our analysis, we find taking $k_{\rm min}=10k_{\rm min,\GW}$ and $k_{\rm max} = 10k_{\rm max,\GW}$ to work well for most examples.} The free nodes are ordered such that $k_{i-1}<k_i<k_{i+1}$, using a bijective ordering transformation that maps a set of variables in $[0,1]$ to an ordered set of variables in the range $[k_{\rm min},k_{\rm max}]$ (see e.g.~\cite{Buscicchio_prior:2019,Handley:2019fll}). Such transformation is characterized by a constant Jacobian, thus if we initially place flat priors on the node locations -- as we do here -- the ordering transform preserves this flat shape of the prior. For the power spectrum amplitudes  at the node positions, we take uniform priors $\log_{10} A_i \in [-8,-1]$, unless otherwise stated.

\smallskip

This free-form reconstruction approach allows the data to determine the shape of the power spectrum by adjusting both the positions and amplitudes of the spline nodes. However, increasing model flexibility risks overfitting, which can be mitigated by selecting the optimal number of nodes using Bayesian evidence -- naturally balancing model complexity against goodness of fit. Building on this idea,  we develop a fully Bayesian framework tailored to this problem. Similar strategies have been employed to reconstruct the primordial curvature perturbations from CMB anisotropies (see e.g.,~\cite{Planck:2018jri,Handley:2019fll} and references therein).

\subsection{Bayesian Model Selection via  Bayesian Evidence}

Since the complexity of the model depends on the number of spline nodes, we employ Bayesian model selection to determine the optimal number of nodes, denoted with $N$. In Bayesian inference, model comparison is performed using the Bayesian evidence (also known as the marginal likelihood),  given by
\begin{equation}
\label{eq:Z_def}
    \mathcal{Z}(\mathcal{M}) = \int \mathcal{L}(d|\theta,\mathcal{M}) \pi(\theta|\mathcal{M}) \, d\theta,
\end{equation}
where $\mathcal{L}(d|\theta,\mathcal{M})$ is the likelihood of the data $d$ given the parameters $\theta$ and model $\mathcal{M}$, and $\pi(\theta|\mathcal{M})$ is the prior distribution of the model parameters~\cite{Trotta:2008qt}. Thus, for each spline model, the parameters $\theta$ over which we integrate are the spline node locations $\log_{10} k_i$, and the corresponding amplitudes $\log_{10}A_i$ of the scalar spectrum. 

\medskip

The Bayesian evidence automatically incorporates a penalty for excessive model complexity, effectively implementing Occam's razor.
In fact, a model with too many spline nodes  has a larger parameter space to integrate over,  diluting the integral and leading to a lower evidence {\it unless} the additional complexity is justified by the data.  By comparing $\mathcal{Z}$ across different values of $N$, we select the models that best balance flexibility and predictive power.  Typically, differences of $\Delta \log \zc \gtrsim 5$ between two models $(1,2)$ suggests that model $(1)$ is decisively favoured over model $(2)$, with $\Delta\log \zc_{(12)} = \log \zc_{(1)} - \log \zc_{(2)} $, while smaller differences ($\Delta \log \zc \lesssim 1$) are considered to be inconclusive in terms of Jeffrey's scale~\cite{Trotta:2008qt}.

\medskip

The posterior distribution for the spline parameters can be highly complex, featuring multiple local maxima,  as well as strong parameter degeneracies. Traditional sampling methods such as Markov Chain Monte Carlo (MCMC) struggle in such cases due to the difficulty of efficiently exploring the high-dimensional space. Additionally, local optimization methods are inadequate because they can become trapped in local extrema. For the same reasons, we opt not to use Akaike information criterion or other frequentist criteria for model comparison. To overcome these challenges, we employ nested sampling~\cite{SkillingNS}, a  particularly well-suited
approach for computing Bayesian evidence while also generating posterior samples as a byproduct. 
\medskip

Nested sampling generates an estimate of the integral in \cref{eq:Z_def}  by reducing the integration space,  focusing on the equivalent one-dimensional integral 
\begin{align}
    \label{eq:Z_ns}
    \zc = \int \mathcal{L}(X) \,dX\,,
\end{align}
where $X$ is a  prior volume corresponding to the likelihood $\bar{\lc}$, defined as 
\begin{align}
    X(\bar{\lc}) = \int_{\lc(\theta) > \bar{\lc} } \pi(\theta)\,d\theta\,.
\end{align}
Hence, $\bar{\lc}$ can be thought of as the enclosing likelihood of cumulative  prior mass $X$ which encompasses the region where the likelihood is larger than a threshold $\bar{\lc}$. 
The core algorithm begins with a set of points  drawn from the prior, called live points. At each step of the algorithm, the live point with the lowest likelihood is replaced by sampling from the prior, subject to the constraint that the likelihood at the new point must be larger than the current lowest likelihood value, and a statistical estimate of the prior mass $X$ is calculated. At the end of this process, we are left with a set of samples $\{\lc,X\}$ which can be used to estimate the evidence as well as approximate the posterior distribution of the parameters $\theta$~\cite{2009BMC,Ashton:2022grj}. This method has several advantages over MCMC: in addition to providing an estimate of the evidence, it can efficiently handle multi-modal posteriors, and  a complex posterior geometry. We implement nested sampling using the publicly available libraries \texttt{PolyChord}~\cite{Handley_2015,Handley:2015fda} and  \texttt{Nautilus}~\cite{Lange:2023ydq}.

\subsection{Functional posteriors}
After obtaining posterior samples for the node parameters, we visualize the reconstructed power spectrum using functional posterior plotting. For a given model, instead of plotting individual samples or  posterior distributions for the spline node locations and amplitudes --  which can be difficult to interpret -- we use the distribution of the sampled node locations and amplitudes to estimate the probability distribution of $\mP_\zeta$ and $\Omega_{\GW}$ at each wavenumber (or frequency). As an example, the mean of the power spectrum amplitude at a given value of $k$ can be written as
\begin{align}
    \mathbb{E}[\mP_{\zeta, k} | \mathcal{M}] = \int d\theta\,  \mP_{\zeta,k}(\theta,\mathcal{M}) p(\theta|d,\mathcal{M})\,,
\end{align}
where  $\mP_{\zeta,k}(\theta,\mathcal{M})$ denotes the power spectrum amplitude at a wavenumber $k$ for the node locations and amplitudes 
under the spline model $\mathcal{M}$, while  $p(\theta|d,\mathcal{M})$ is the posterior distribution of the spline parameters under the model $\mathcal{M}$.

\smallskip

In the spirit of Bayesian inference, we further marginalize over all the models with different number of nodes, using the evidence ratios as the weights for the marginalization. Following such procedure, the above estimate of the mean -- taken over all the models with different number of nodes $n$ -- can then be expressed as
\begin{align}
    \mathbb{E}[\mP_{\zeta, k}] = \frac{\sum_n \mathbb{E}[\mP_{\zeta, k}|\mathcal{M}_n]\zc_n\,}{\sum_n \zc_n}\,.
\end{align}
Similarly, for each frequency bin in momentum $k$ (or frequency $f$)-space, we can compute other summary statistics such as the median and credible intervals for $\mP_\zeta$ and $\Omega_{\GW}$ from the posterior samples.

\section{Results}
\label{sec:results}

We present results for selected representative cases, showing the functional posterior distributions of $\mP_\zeta$ and $\Omega_{\mathrm{GW}}$ when marginalised over models with varying numbers of spline nodes. We also include a plot of the Bayes factor $\log \mathcal{Z}$ as a function of the number of nodes used in the reconstruction of $\mP_\zeta$.
 Unless  otherwise stated, for each of the $\Omega_{\rm GW}$ spectra  considered we generate $1\sigma$ error bars for the observed $\Omega_{\rm GW}$ values at 50 logarithmically spaced frequencies in $5\times 10^{-5} \,\mathrm{[Hz]}< f < 10^{-2}\,\mathrm{[Hz]}$ with 
\begin{align}
\label{eq:DOGW}
    \Delta \Omega_{\rm GW}(f) = \Omega_{\rm GW ,obs}(f) \left[0.1 + 0.05(\log f/f_*)^2  \right]\,.
\end{align}
As an example, we  choose the pivot frequency
$f_{*} = 10^{-3} \,\mathrm{Hz}$ as  representative for LISA, with low uncertainties around the peak sensitivity $f_*$ and increasing ones towards the two ends of LISA's sensitivity range (see e.g. \cite{LISA:2024hlh}). We then use the result  to create a Gaussian likelihood for the $\Omega_{\rm GW}$ as an input for the nested sampler, i.e.
\begin{align}
    \label{eq:Gaussian_likelihood}
    -2\ln \mathcal{L} = \sum_{f} \rm \left(\frac{\Omega_{\GW,\rm model} - {\Omega}_{\GW,\mathrm{obs}}}{\Delta \Omega_{\rm GW}} \right)^2 \,.
\end{align}
In our analysis, we  assume  that such $\Omega_{\GW}$ spectrum is  reconstructed from detector data -- e.g.~using codes such as \texttt{SGWBinner}~\cite{Caprini:2019pxz,Flauger:2020qyi} in the case of LISA {(see also \cite{Hartwig:2023pft,Knapp:2025upw,Santini:2025iuj})} -- thus providing   datasets consisting of the observed values of $\Omega_{\GW}$ and their associated uncertainties $\Delta\Omega_{\GW}$, over  certain frequency ranges specific to the instruments.\footnote{This assumption helps us to focus our attention entirely on our reconstruction method but is not a necessity. This reconstruction method (or any other) can be easily incorporated into existing codes in such a manner that eliminates the need for a separate initial inference step to obtain $\Omega_{\GW},\,\Delta\Omega_{\GW}$. {However, the two-step procedure has the advantage that the reconstruction method need not be applied repeatedly to the full dataset, but only to the $\Omega_{\rm GW}$ data, making it computationally more feasible.}}   

While the spectra and error bars are highly idealised, and artificially generated  with LISA as  representative experiment to keep in mind, the methods we present  do not rely on this being the case. They are applicable for any general form for $\Omega_{\GW}$ and any GW experiment. In particular, one may consider more realistic realizations of the $\Omega_{\GW}$ data simulated using the LISA noise model, without  changing our method.\footnote{{In \cref{app:noisy_realisations} we also present results for noisy realisations of the $\Omega_{\GW}$ data.}} We should note though that the curvature spectrum reconstruction can only be as good as the $\Omega_{\GW}$ reconstruction from detector data.

\subsection{Standard radiation domination}
\label{sec:RD}

\begin{figure}[t]
    \centering
\includegraphics[width=0.9\linewidth]{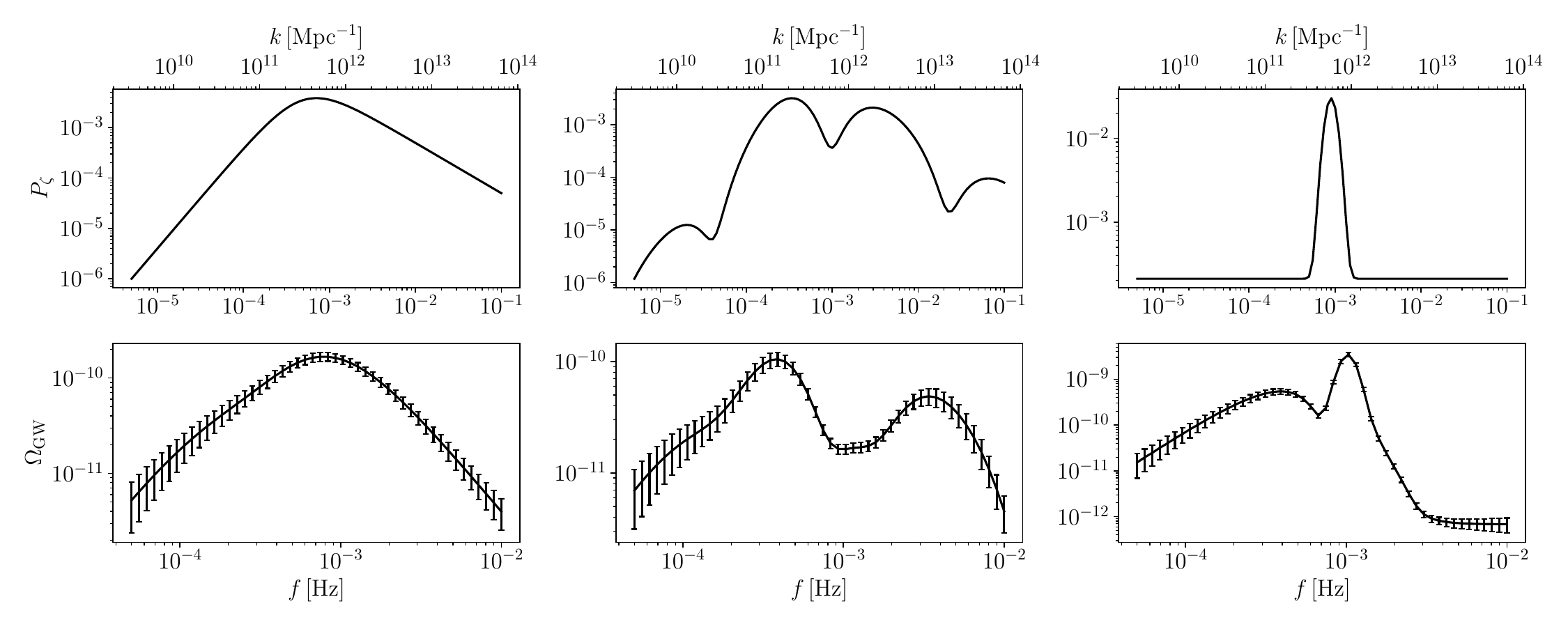}
    \caption{\small Test spectra chosen for power spectrum reconstruction assuming radiation domination. {\it First
    column}: $\mP_\zeta$ and $\Omega_{\rm GW}$
    for the broken power law profile of section \ref{sec:bpl_res}. {\it Second column}: $\mP_\zeta$ and $\Omega_{\rm GW}$
    for the template with oscillatory features of section \ref{sec_temof}. {\it Third column}: $\mP_\zeta$ and $\Omega_{\rm GW}$
    for the peaked profile of section \ref{sec_shapeak}. For each column in the second row, we also show the corresponding error bars on $\Omega_{\GW}$, as dictated by \cref{eq:DOGW}.}
    \label{fig:RD_input}
\end{figure}
We start  discussing our results in
the framework of  standard cosmological expansion, in
which radiation domination directly follows the inflationary epoch. In this case, the transfer function is given by eq.~\eqref{eq:TRD} and $\Omega_{\mathrm{GW},\mathrm{rh}}$ in \cref{eq:Omega_rh_RD} is understood to be evaluated at the time when the source terms have sufficiently decayed, after which the induced tensor perturbations evolve as freely propagating gravitational waves.
 We present reconstruction results for the three distinct scalar spectrum templates shown in \cref{fig:RD_input}. In each case, the scalar spectral shape is held fixed (upper panel), and we attempt to reconstruct it from the  induced gravitational wave signal
 it gives rise to (lower panel). In appendix \ref{app_pta} we test
 our method against recent pulsar timing array measurements.

\subsubsection{Broken power law}
\label{sec:bpl_res}

We start considering a broken power law  (BPL) template shown in the left panel of \cref{fig:RD_input}  (see e.g.  
\cite{Caprini:2024hue,LISACosmologyWorkingGroup:2024hsc,Marriott-Best:2024anh}):
\begin{align}
\label{eq:bpl}
\mP_{\zeta}(f) = A  \left(\frac{f}{f_*}\right)^{n_{\rm IR}}  \left(1 + \left(\frac{f}{f_*}\right)^{\sigma} \right)^{\frac{n_{\rm UV} - n_{\rm IR}}{\sigma}} \,,
\end{align}
where $A$ is the amplitude, $f_*$ a reference frequency,
$n_{\rm IR,UV}$ the spectral indexes in the infrared and ultraviolet
part of the spectrum, while $\sigma$ a parameter
controlling the smoothness of the transition.

\smallskip

To start with,
we fix
the parameter values of  \cref{eq:bpl} as
\begin{equation}
A = 10^{-2},\,\,\, f_* = 5\times 10^{-4}\,\mathrm{Hz} ,\,\,\,n_{\rm IR}=2,\,\,\, n_{\rm UV}=-1,\,\,\,\sigma =2,
\label{eq_pabplc1}
\end{equation}
to ensure that the peak of the induced GW spectrum lies in the middle of the LISA band.

\begin{figure}
    \centering
    \includegraphics[width=0.85\linewidth]{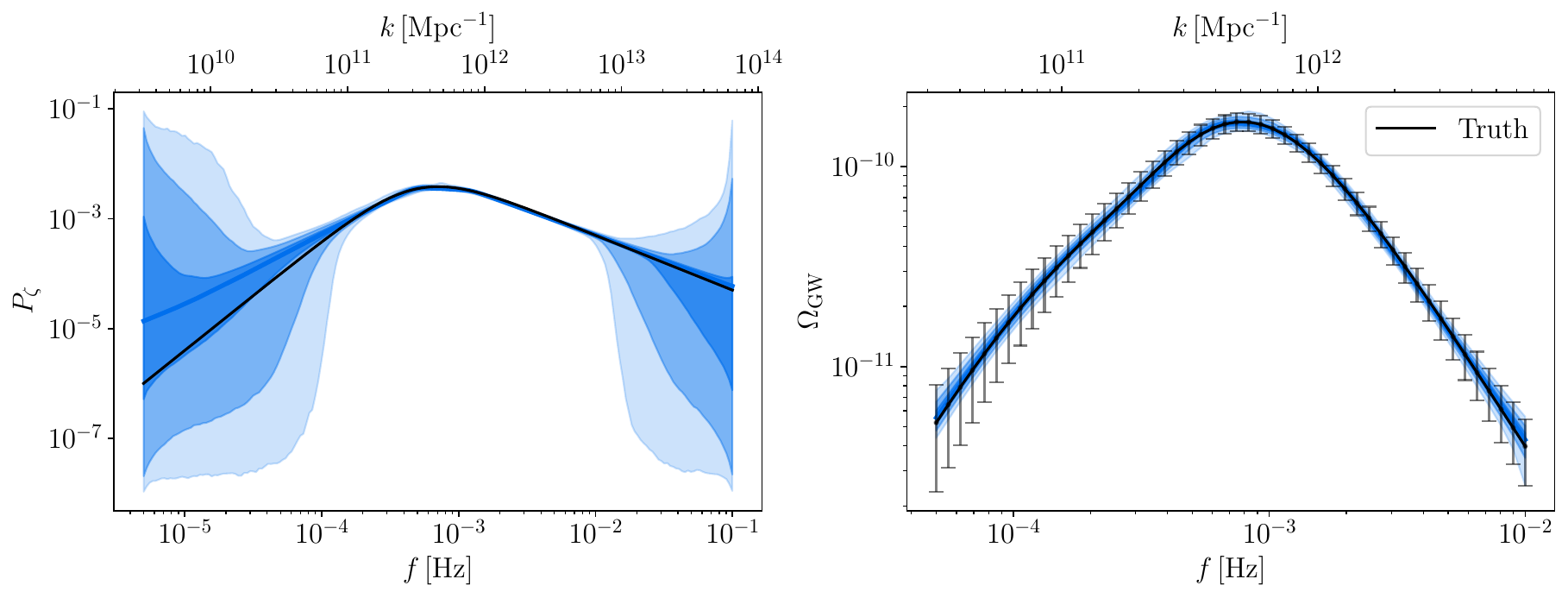}
    \includegraphics[width=0.45\linewidth]{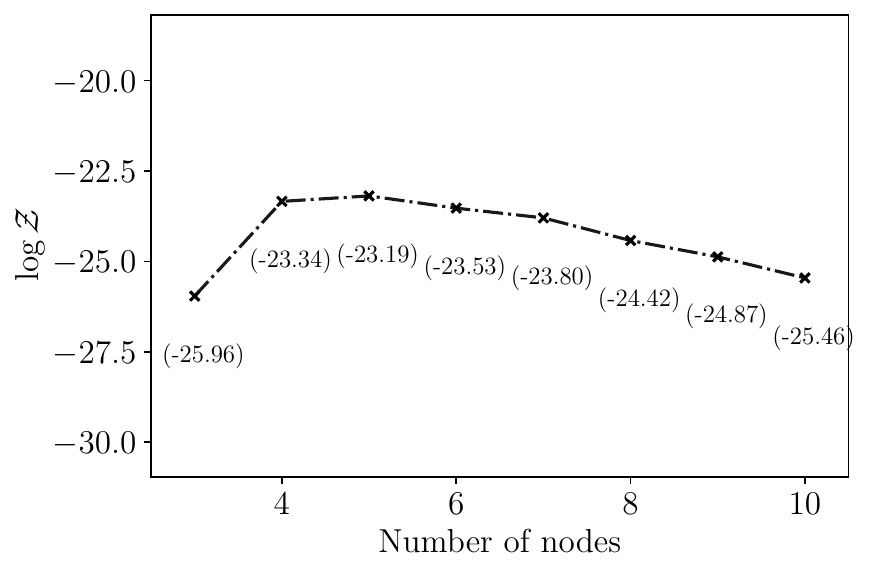}
    \caption{\small {\it Top}: The reconstructed $\mP_\zeta$ and $\Omega_{\GW}$ for the BPL model of section \ref{sec:bpl_res}, with parameters as in eq \eqref{eq_pabplc1}, marginalised over the models with different number of nodes. The different shaded regions correspond to $68\%, 95\%$ and $99.7\%$ credible intervals for $\mP_\zeta$ and $\Omega_{\GW}$. {\it Bottom}: The evidence $\log \mathcal{Z}$ as a function of the number of nodes.} 
    \label{fig:bpl_rd}
\end{figure}

Our results  are shown in \cref{fig:bpl_rd}. {The Bayesian evidence here favours the $N=4,5$ model the most and decreases as we increase the number of nodes, suggesting that the shape of the spectrum can be captured sufficiently well with 4 and 5 nodes.\footnote{{Nested sampling produces a statistical estimate of $\log \mathcal{Z}$, thus these numbers in  \cref{fig:bpl_rd} come with their associated uncertainties. When running \texttt{Nautilus} to sample the posterior and compute the evidence, we choose settings to ensure that the resulting uncertainty is small $\Delta \log \mathcal{Z} < 0.1$ (see \cite{Lange:2023ydq} and also \texttt{Nautilus} \href{https://nautilus-sampler.readthedocs.io/en/latest/discussion/faqs.html}{FAQs}) Since this is much smaller compared to the range over which $\log \mathcal{Z}$ varies as a function of the number of nodes, we do not show it in the plots.}}} 

We find that the scalar power spectrum $\mP_\zeta$ can be accurately reconstructed near the peak, while larger uncertainties appear at the edges, reflecting the limitations of the input data. In particular, the constraints on the power spectrum in the infrared (IR) are almost entirely prior-dominated. This is due to two key reasons: \textit{i)} IR modes far from the peak contribute negligibly to the SIGW spectrum around the peak, and \textit{ii)} the IR tail of the SIGW spectrum generated during radiation domination exhibits a universal $k^3$ behaviour, up to logarithmic corrections (see e.g.~\cite{Caprini:2009fx,Cai:2019cdl,Yuan:2019wwo,Domenech:2021ztg,LISACosmologyWorkingGroup:2024hsc,Domenech:2025bvr}). Nevertheless,
it is known that in single field inflation $\mP_\zeta$
can not normally growth steeper than $k^4$ in the infrared \cite{Byrnes:2018txb,Ozsoy:2019lyy} (but
see e.g.~\cite{Carrilho:2019oqg,Tasinato:2020vdk,Tasinato:2023ukp,Cielo:2024poz} for exceptions).

\medskip

\begin{figure}
    \centering
    \includegraphics[width=0.85\linewidth]{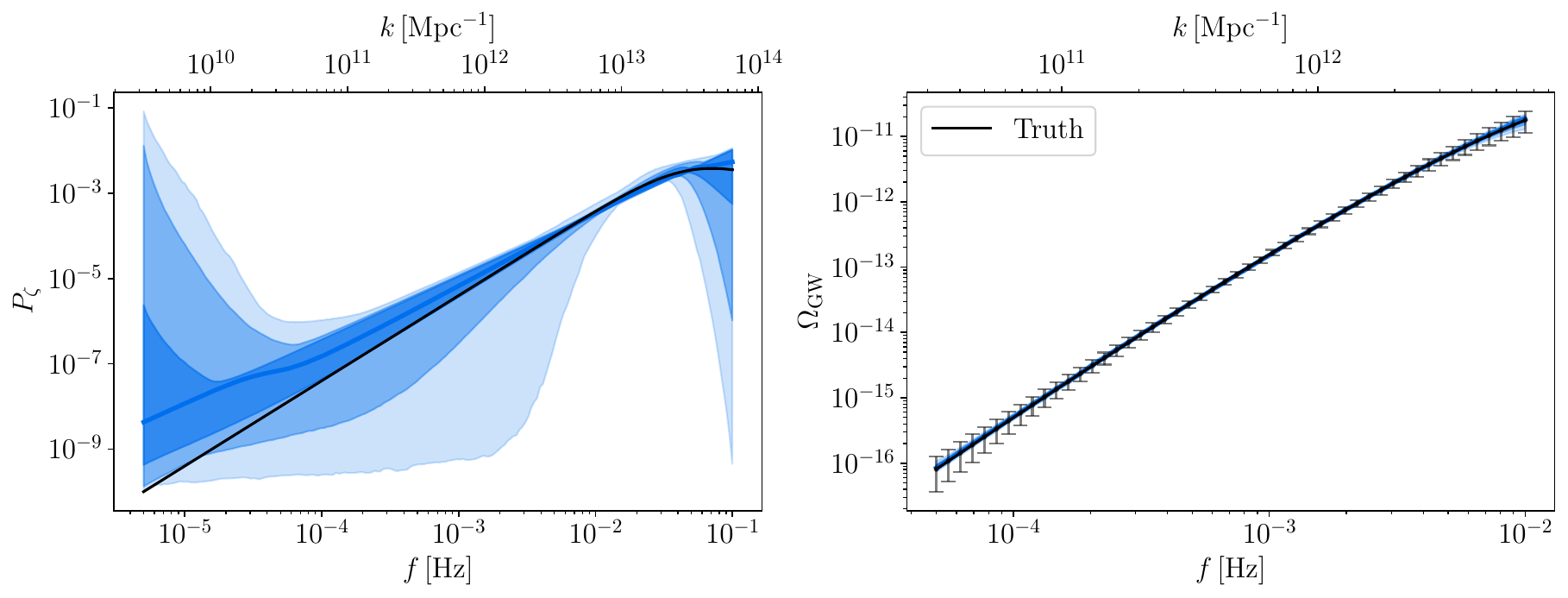}
    \includegraphics[width=0.85\linewidth]{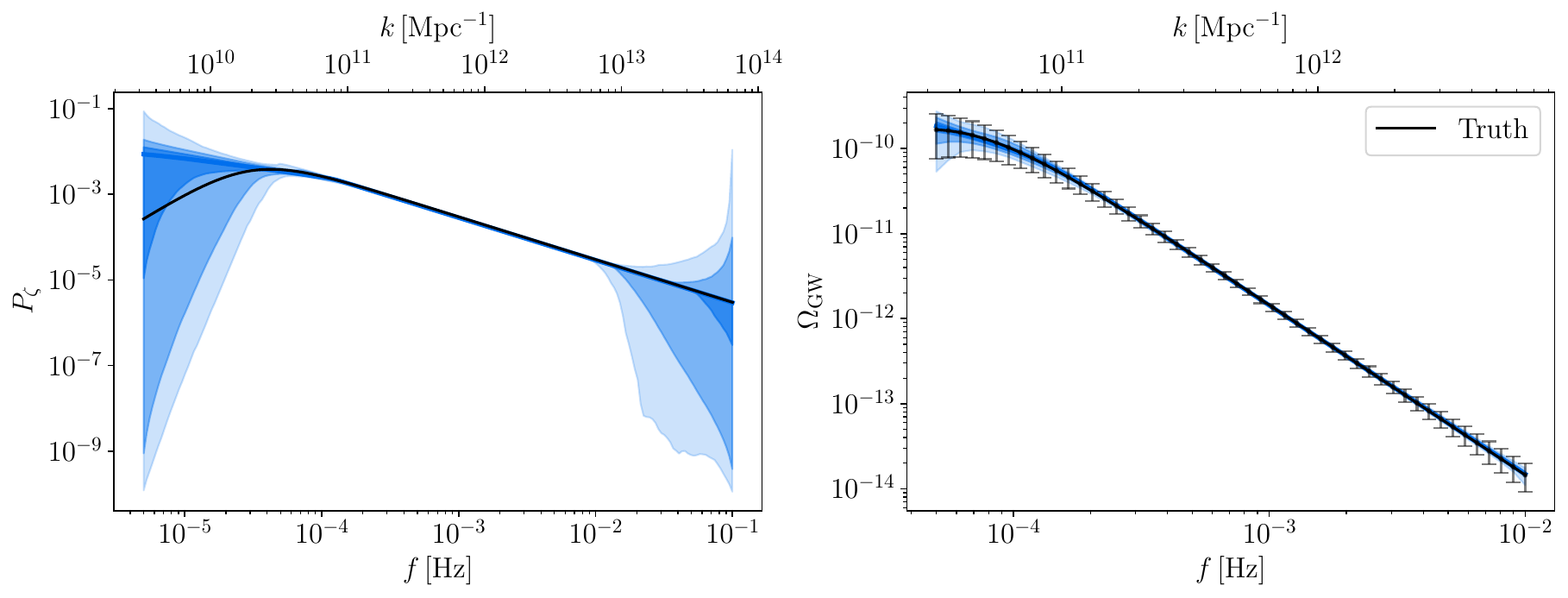}
    \caption{\small Reconstruction of the BPL power spectrum using if only the IR (top) and UV (bottom) part of the spectrum were observed by the experiment. The different shaded regions correspond to $68\%, 95\%$ and $99.7\%$ credible intervals for $\mP_\zeta$ and $\Omega_{\GW}$.}
    \label{fig:BPL_UV_IR}
\end{figure}

As a second case, we  assume that only the UV or IR portion of the spectrum is detected by the experiment. We use the same parameters as the original BPL example, except for $f_* = 3\times 10^{-5},5\times 10^{-2}$ Hz, for the UV and IR cases respectively. The corresponding results are shown in \cref{fig:BPL_UV_IR}. In the IR, the universal \( k^3 \) scaling of the \( \Omega_{\mathrm{GW}} \) spectrum far from the peak implies that \( \mP_\zeta \) cannot be reliably reconstructed in the tails, where the constraints are almost entirely prior-dominated. In contrast, the UV tail of the \( \Omega_{\mathrm{GW}} \) spectrum is sensitive to the UV tilt of the curvature power spectrum, hence enabling its reconstruction\footnote{We noticed a similar behaviour for  additional radiation-dominated examples we explored.}.
These results can be understood analytically by noting that if \( n_{\mathrm{IR}} \) and \( n_{\mathrm{UV}} \) denote the IR and UV tilts of the curvature power spectrum, then the induced \( \Omega_{\mathrm{GW}} \) spectrum exhibits the following asymptotic behavior (see e.g.~\cite{Domenech:2021ztg}):
\begin{align}
\Omega_{\mathrm{GW}}(k)\;\propto\;
\begin{cases}
\displaystyle
\biggl(\frac{k}{k_{\rm p}}\biggr)^{3}, & k \ll k_{\rm p},\\[1ex]
\displaystyle
\biggl(\frac{k}{k_{\rm p}}\biggr)^{-\Delta}, & k \gg k_{\rm p},
\end{cases}
\qquad
\text{with}
\quad
\Delta =
\begin{cases}
2\,n_{\mathrm{UV}}, & 0 < n_{\mathrm{UV}} < 4,\\[0.5ex]
4 + n_{\mathrm{UV}},   & n_{\mathrm{UV}} > 4,
\end{cases}
\label{eq:approx_bpl_RD}
\end{align}
where $k_{\rm p}$ is the peak scale. It is clear from the above equations that for such BPL profiles of the curvature power spectrum, the IR tilt plays no role in the shape of the SIGW spectrum whereas the UV tilt does. This is very much reflected in the marginalised posteriors of the curvature power spectrum for the all the examples seen previously.  

\subsubsection{A template with oscillatory features}
\label{sec_temof}

\begin{figure}
    \centering
    \includegraphics[width=0.85\linewidth]{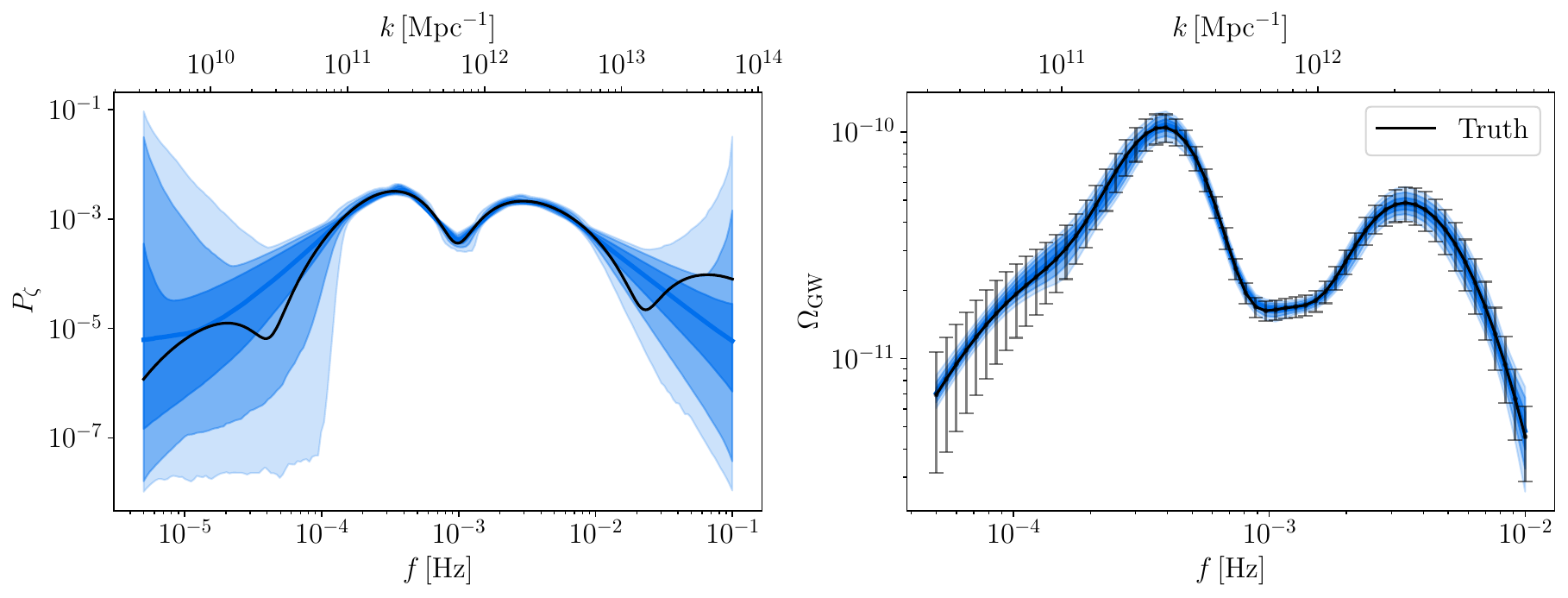}
    \includegraphics[width=0.45\linewidth]{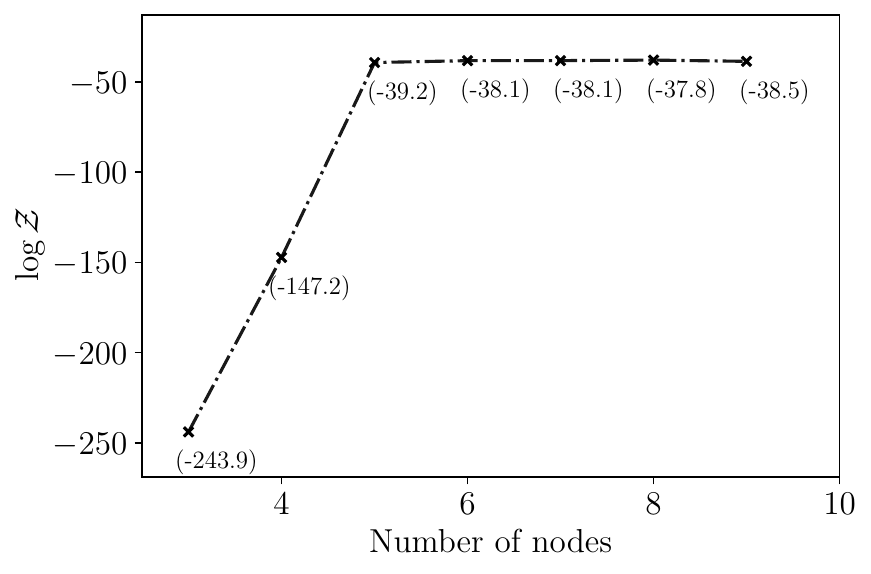}
    
    \caption{\small {\it Top}: The reconstructed $\mP_\zeta$ and $\Omega_{\GW}$ for the oscillatory model of eq \eqref{eq:osc}, marginalised over the models with different number of nodes. See section \ref{sec_temof}. The different shaded regions correspond to $68\%, 95\%$ and $99.7\%$ credible intervals for $\mP_\zeta$ and $\Omega_{\GW}$. {\it Bottom}: The evidence $\log \mathcal{Z}$ as a function of the number of nodes.}
    \label{fig:osc_rd}
\end{figure}

As a second representative template for the curvature 
power spectrum, we modify the BPL profile \eqref{eq:bpl}
including oscillatory features  (middle  column, \cref{fig:RD_input}. See e.g.  \cite{Fumagalli:2021cel,Witkowski:2021raz,Braglia:2020taf,LISACosmologyWorkingGroup:2025vdz} for physical motivations)
\begin{align}
    \label{eq:osc}
    \mP_{\zeta}(f) = A \cdot \left(\frac{f}{f_*}\right)^{n_{\rm IR}} \cdot \left(1 + \left(\frac{f}{f_*}\right)^{\sigma} \right)^{\frac{n_{\rm UV} - n_{\rm IR}}{\sigma}} \left(1 + B \cos \left( (\ln f /C)^2 \right) \right)
\end{align}
The new parameters $A,\,B,\,C$ control the oscillations.
We fix their values  $A=10^{-3}, B=15, C=2.5$ while the other parameters of  \cref{eq:osc} have the same values as their BPL counterparts in
\cref{eq_pabplc1}. The corresponding results are represented in \cref{fig:osc_rd}.  The  plot indicates that our Bayesian method is able to recover the shape of the oscillations quite well.  Once again, the constraints in the tails are much weaker, especially in the IR since the IR part of the $\mP_{\zeta}$ spectrum contributes very little to the SIGW spectrum for peaked $\mP_\zeta$ spectra.

\subsubsection{A sharply peaked spectrum}
\label{sec_shapeak}

\begin{figure}
    \centering
    \includegraphics[width=0.85\linewidth]{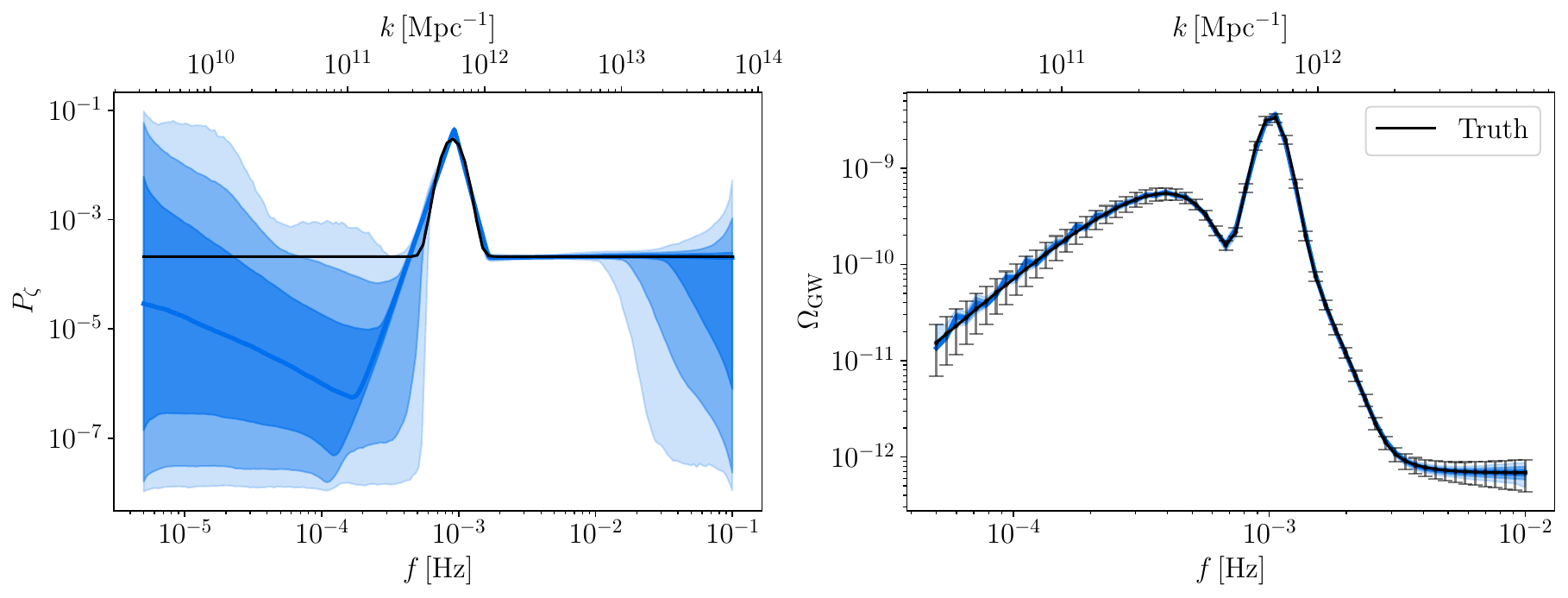}
    \includegraphics[width=0.45\linewidth]{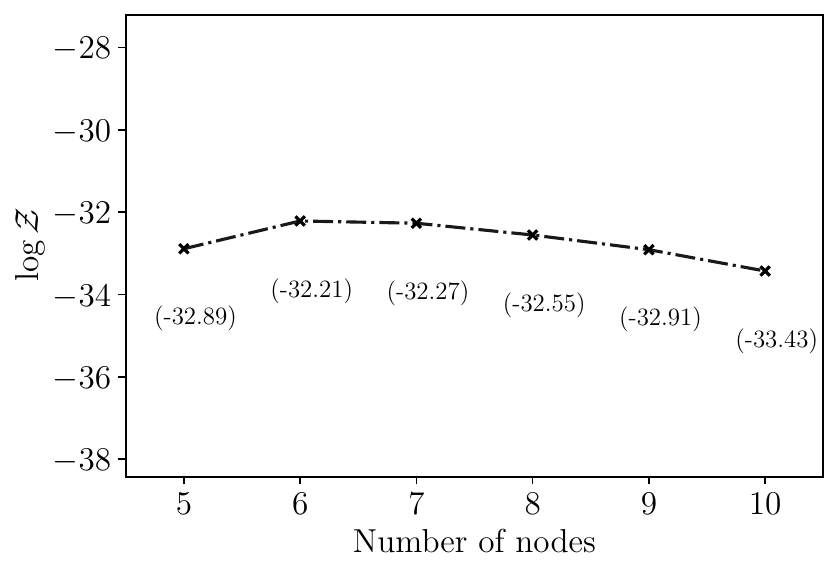}
    \caption{\small {\it Top}: The reconstructed $\mP_\zeta$ and $\Omega_{\GW}$ for the peaked model of eq \eqref{eq:peaked}, marginalised over the models with different number of nodes. See section \ref{sec_shapeak}. The different shaded regions correspond to $68\%, 95\%$ and $99.7\%$ credible intervals for $\mP_\zeta$ and $\Omega_{\GW}$. {\it Bottom}: The evidence $\log \mathcal{Z}$ as a function of the number of nodes.}    \label{fig:peaked_rd}
\end{figure}

We next try to reconstruct
a curvature spectrum characterized by a sharp feature, described by
a modified log-normal profile~\footnote{Log-normal
scalar profiles have been much explored in the recent literature. See e.g. \cite{LISACosmologyWorkingGroup:2025vdz} for 
motivations and examples.} (right column, \cref{fig:RD_input}):
\label{eq:peaked}
\begin{align}
    \mP_\zeta(f) = A \cdot \left( B + \exp \left( -\frac{1}{2} \left( \frac{\ln (f / f^*)}{C} \right)^2 \right) \right)
\end{align}
We select the parameters as $A=3\times 10^{-2},B=7\times 10^{-3}, f_* = 9\times 10^{-4}\,\mathrm{Hz},\, C=0.15$, so to ensure that the corresponding GW spectrum is enhanced in the LISA band. The results for this example are plotted in \cref{fig:peaked_rd}. Notice
that the position and features of the peak of the curvature spectrum are reconstructed extremely well since the resonant peak and dips in the SIGW spectrum are highly dependent on the location of the peak in the $\mP_{\zeta}$ spectrum.

\subsubsection{Non-detection}
\label{sec_nond}
\begin{figure} 
    \centering
    \includegraphics[width=0.85\linewidth]{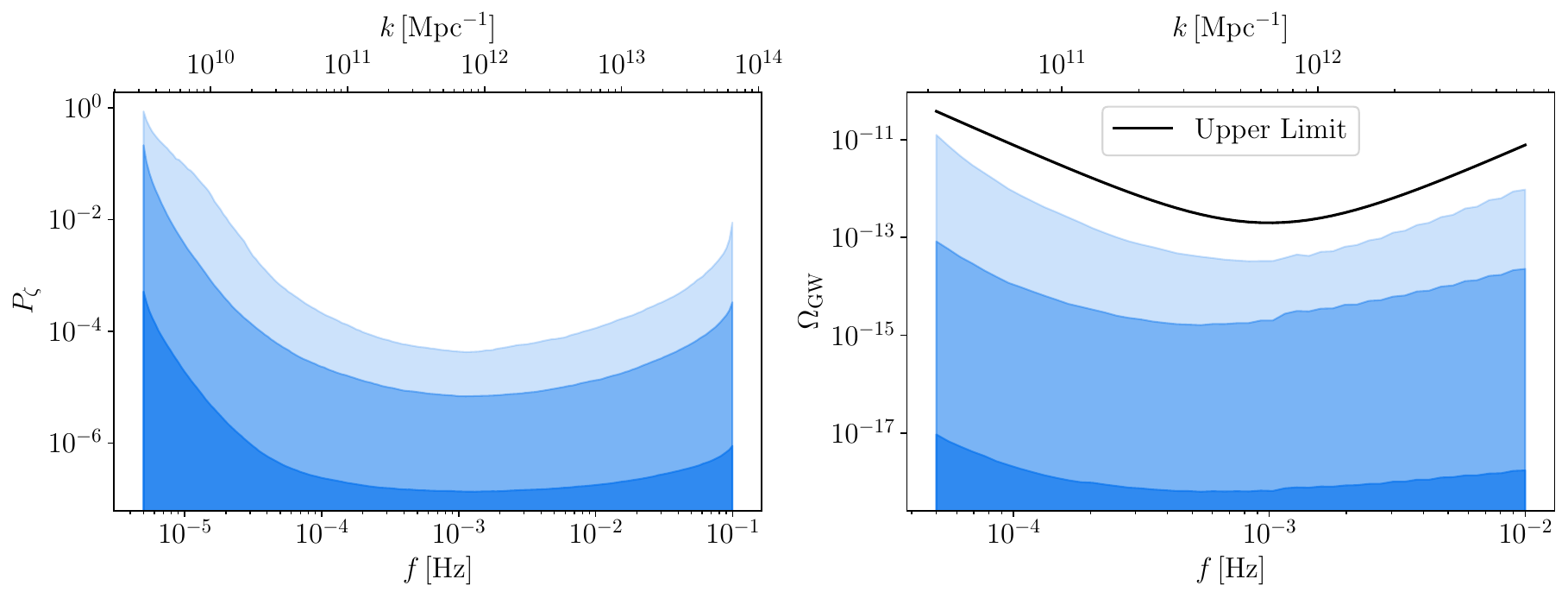}
    \caption{\small  The upper limits for $\mP_\zeta$ reconstructed using 4 nodes from the upper limit on $\Omega_{\GW}$. The boundaries of the different shaded regions correspond to $68\%, 95\%$ and $99.7\%$ upper limits. See section \ref{sec_nond}.}
    \label{fig:UL_RD}
\end{figure}

Finally, we present results for the case of a non-detection in \cref{fig:UL_RD}, where we only have upper limits on the $\Omega_{\GW}$ over a given frequency range. For a given detector, the upper limits on $\Omega_{\GW}$ would be the strongest around the peak sensitivity of the detector and weaker towards the ends, which is reflected in our mock upper limits shown in \cref{fig:UL_RD}. Since there is no detection in this case, we fix the number of nodes to four for simplicity although one can again obtain upper limits marginalised over models with different number of spline nodes as done in the previous examples. Our results demonstrate that this reconstruction method can also be applied to obtain upper limits on $\mP_\zeta$ over a given frequency/wavenumber range which can complement the limits on $\mP_\zeta$ obtained from PBH constraints. Given the exponential sensitivity of the PBH abundance to the perturbation amplitude~\cite{LISACosmologyWorkingGroup:2023njw}, such limits can be very powerful in constraining the fraction of dark matter in PBH.

\subsection{Early matter domination-radiation transition}
\label{sec_EMD}

In the previous section we focused on scenarios where  primordial GWs are amplified during radiation domination
upon re-entering the horizon. In this section and the next, we introduce an additional layer of complexity by considering cases where the universe is not necessarily radiation dominated at the time of GW formation. In particular, we consider a cosmological scenario in which inflation is followed by a period of matter domination that transitions quickly to RD, as laid out in  \cref{sec:sigwkernels}. Such an early matter domination (eMD) epoch is characterised by an equation of state parameter $w = 0$, and we  impose the relation $c_{\mathrm{s}}^2=w$ for the fluid sound speed \cite{Assadullahi:2009nf}. 

Scenarios involving an eMD phase have gained increasing attention in recent years due to their potential to enhance the amplitude of induced GWs, bringing the resulting GW signal within the sensitivity range of future detectors such as LISA, offering a potential observational probe of the thermal evolution of the universe.

To calculate the induced GWs, we impose an abrupt cutoff on the power spectrum of curvature perturbations at $k_{\mathrm{cut}}$ which may be interpreted as the scale corresponding to the onset of eMD, or the scale where density perturbations become non-linear
$k_{\mathrm{cut}} \sim 470 / \eta_{\mathrm{rh}}$ \cite{inomata_enhancement_2019}:
\begin{equation} \label{eq:Ph_cut}
    \mathcal{P}_\zeta(k)=0 \quad \text { for } k>k_{\mathrm{cut}}.
\end{equation}

Our aim is  to evaluate whether our Bayesian reconstruction method remains accurate and well-behaved in the presence of the reheating transition described above. To this end, we adopt the following benchmark parameters:
\begin{equation}
\eta_{\rm rh} = 2500\,\mathrm{s},
\qquad 
k_{\mathrm{cut}} = 0.008 \, \mathrm{s}^{-1}.
\end{equation} 

To further test the reconstruction capabilities of our method, we move beyond the standard assumption of a scale-invariant spectrum and consider two representative templates for $\mathcal{P}_\zeta$: the broken power-law profile of section \ref{sec:bpl_res}, and the narrow  peak profile of section \ref{sec_shapeak} (see \cref{fig:RD_input}).

\begin{figure}
    \centering
    \includegraphics[width=0.85\linewidth]{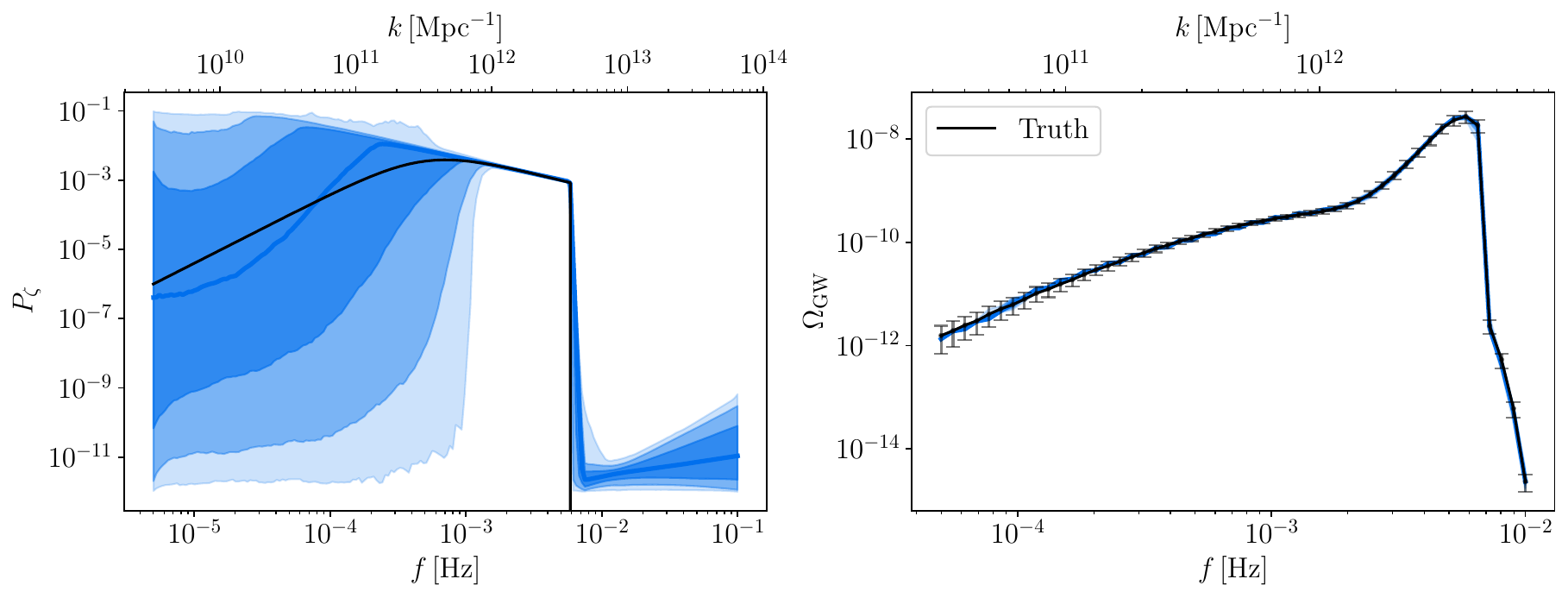}
    \includegraphics[width=0.45\linewidth]{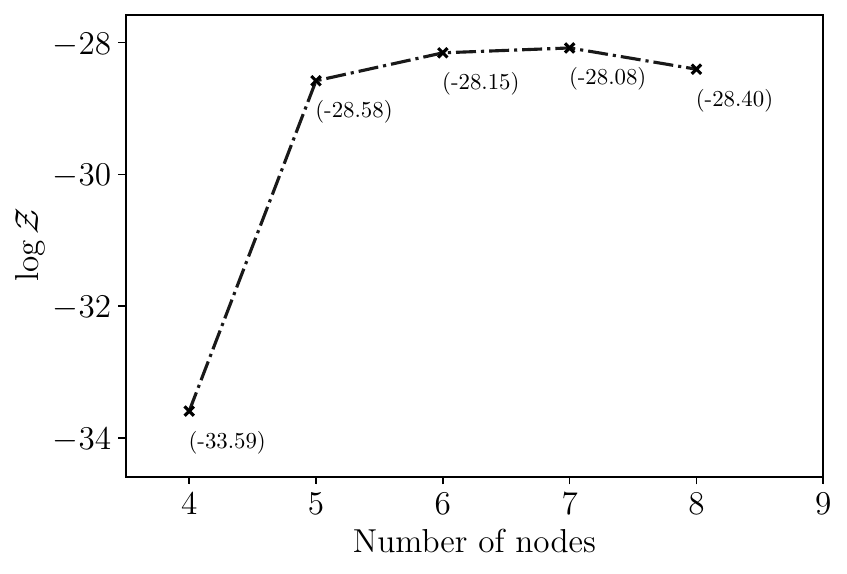}
    \caption{\small {\it Top}: The reconstructed $\mP_\zeta$ and $\Omega_{\GW}$ for the BPL model, assuming early matter domination, marginalised over the models with different number of nodes. See  \cref{sec_EMD}. The different shaded regions correspond to $68\%, 95\%$ and $99.7\%$ credible intervals for $\mP_\zeta$ and $\Omega_{\GW}$. {\it Bottom}: The evidence $\log \mathcal{Z}$ as a function of the number of nodes.}
    \label{fig:bpl_mdrd}
\end{figure}

\begin{figure}
    \centering
    \includegraphics[width=0.85\linewidth]{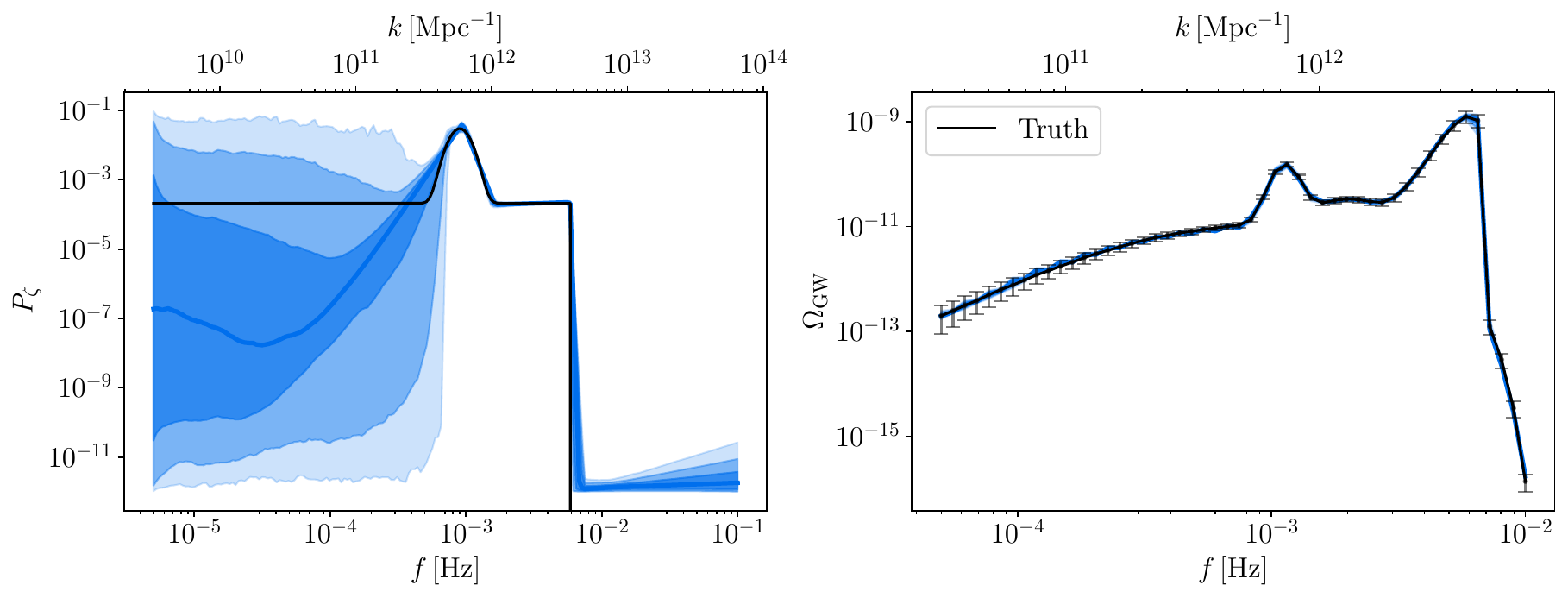}
    \includegraphics[width=0.45\linewidth]{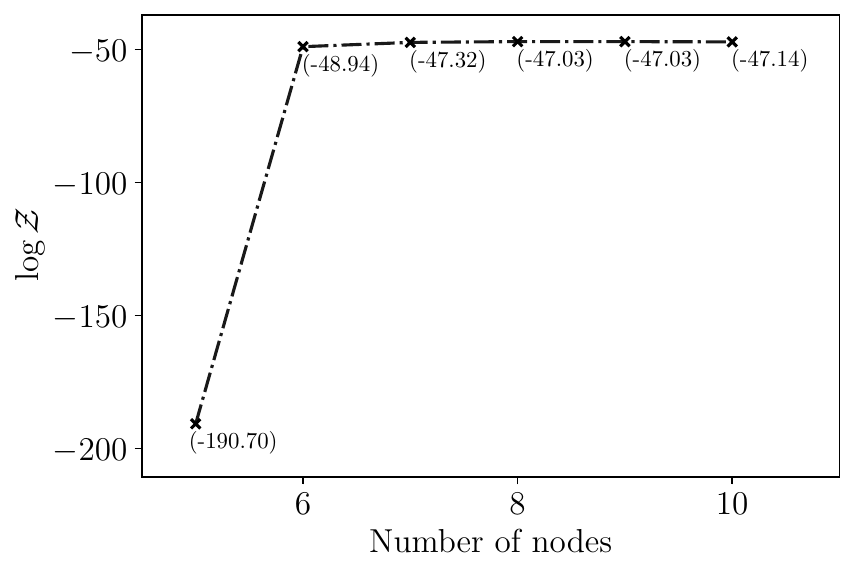}
    \caption{\small {\it Top}: The reconstructed $\mP_\zeta$ and $\Omega_{\GW}$ for the peaked model in eMD, marginalised over the models with different number of nodes. {\it Bottom}: The evidence $\log \mathcal{Z}$ as a function of the number of nodes.}
    \label{fig:peaked_mdrd}
\end{figure}

The results for the BPL model with an eMD-to-RD transition are shown in  \cref{fig:bpl_mdrd}. The reconstruction succesfully recovers the shape and amplitude of the power spectrum near the peak and around the cutoff. Larger uncertainties appear at lower frequencies, which  are largely prior-dominated, and reflect the limited constraining power of the stochastic GW background  in the infrared regime. At high frequencies, uncertainties arise from the shape of the resonance contribution to the GW kernel, which is sharply peaked around the transition,  and results quite sensitive to the choice of reheating time.

Considering the narrow-peaked spectrum of  \cref{fig:peaked_mdrd}, the reconstruction recovers the location and amplitude of the peak with good accuracy, hence
demonstrating its sensitivity to sharply localised features in the power spectrum. As with the BPL case of  \cref{sec:bpl_res}, we find  uncertainties in the reconstruction of the infrared part, where the induced GW spectrum becomes less constraining and the reconstruction becomes increasingly prior dominated.

\begin{figure}[ht]
    \centering
    \includegraphics[width=0.85\linewidth]{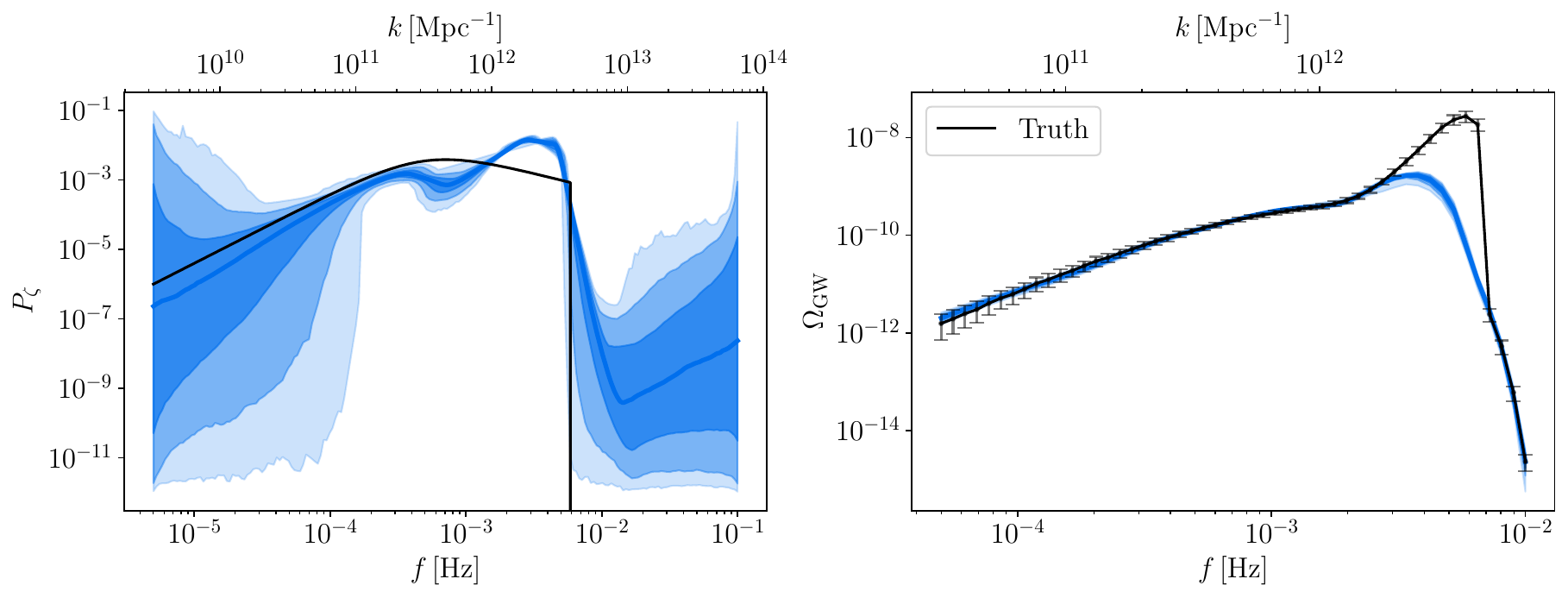}
    \includegraphics[width=0.45\linewidth]{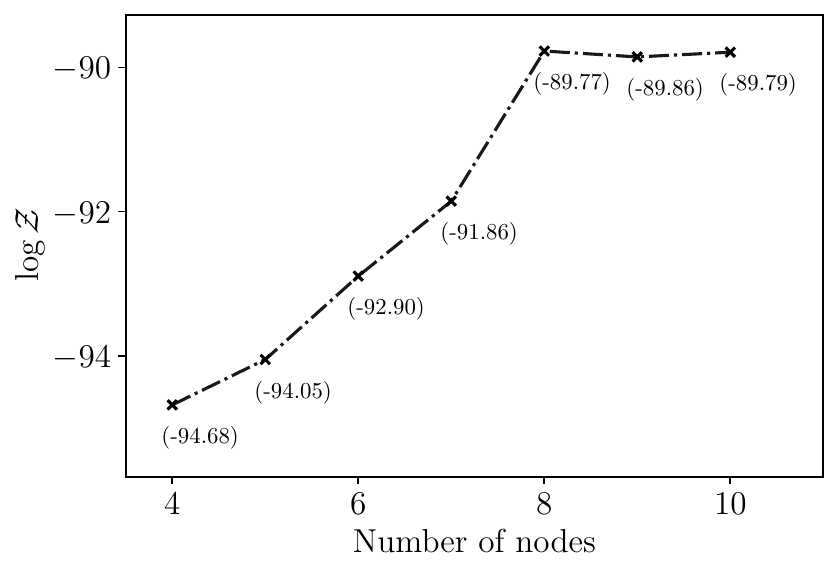}
    \caption{\small {\it Top}: The reconstructed $\mP_\zeta$ and $\Omega_{\GW}$ for the BPL model assuming eMD domination,  with a run assuming RD for the same GWB data, and marginalised over the models with different number of nodes. The different shaded regions correspond to $68\%, 95\%$ and $99.7\%$ credible intervals for $\mP_\zeta$ and $\Omega_{\GW}$. {\it Bottom}: The evidence $\log \mathcal{Z}$ as a function of the number of nodes.}
    \label{fig:bpl_rd_mdrd}
\end{figure}

Proceeding with additional tests,  we  perform for comparison a reconstruction assuming a pure radiation-dominated background using GW data generated under an early matter-dominated to radiation-dominated (eMD-RD) scenario for the  BPL model. As shown in  \cref{fig:bpl_rd_mdrd}, a resonance peak arising from the sudden reheating transition acts as a distinguishing feature between the two scenarios, with the Bayes factor decisively favouring the eMD-RD model. This indicates that, in principle, GW observations can be used to infer features of the transition.  However, the enhancement in induced GWs is sensitive to the dynamics of the reheating process. While a sudden transition can lead to a pronounced enhancement, more gradual reheating, such as in models with constant decay rates \cite{Inomata_gradual_2019} can suppress the signal, potentially weakening the apparent model preference. However, if the decay rate evolves in time, such that the field decays much more rapidly than the Hubble expansion rate at the time of transition, the induced GWs can still be significantly enhanced \cite{pearce2024gravitationalwavesignalsearly}. Thus, whether a bias in model reconstruction persists depends on the specific details of the transition dynamics, and further study is needed to assess the robustness of these conclusions across various reheating scenarios.

Additionally, it would be particularly interesting to explore whether the time dependence of the gravitational potential and its decay rate could be reconstructed from the GW signal itself. If this is feasible, it could provide insights into the underlying physics and thermal history, potentially allowing us to pinpoint key features such as (a) the onset of the eMD era, (b) its duration, and (c) the timescale of the transition to RD. Recent studies have begun to map the impact of these factors on $\Omega_\text{GW}$ \cite{pearce2024gravitationalwavesignalsearly, Inomata_gradual_2019, inomata_enhancement_2019, pearce2025usinggravitationalwavesignals}, and using the reconstruction methods proposed in this work, it might even be possible to directly infer this information from real signals, such as those detectable by LISA, offering a direct observational probe into the thermal history of the early universe, and possibly also the QCD transition (see e.g. 
\cite{Byrnes:2018clq,Franciolini:2023wjm}). We leave these considerations for future study.

\subsection{General equation of state}
\label{sec:gen_w}

We now consider the more general case in which  the equation of state of the dominant fluid at the time of horizon re-entry of the scalar modes can in principle differ from that of matter or  radiation. 

The transfer function $\mathcal{T}_w$  is given by \cref{eq:Tgenw} \cite{Domenech:2021ztg}. 
Our aim in this case  is 
to use GW data not only for reconstructing the momentum shape of the primordial power spectrum, but also the equation of state $w$ (we fix the sound speed such that $w=c_s^2$, i.e. treating this component as a perfect fluid). We make use of a modified version of the public code \texttt{SIGWFast}~\cite{Witkowski:2022mtg} to obtain the $\Omega_{\GW}(f)$ spectrum given an input curvature power spectrum, equation of state $w$ and transition/reheating scale $f_{\rm rh}$. \sigwfast~calculates $\Omega_{\GW}(f)$ under the assumptions stated above (we take $f_{\rm rh} = 10^{-5}\rm
\,Hz$).\footnote{Note that the code can only compute $\Omega_{\GW}(f)$ for $f>f_{\rm rh}$.} We do not sample $f_{\rm rh}$ since it appears as an overall scaling  factor in $\Omega_{\GW}(f)$, thus it is completely degenerate with the power spectrum amplitude. We leave an analysis of how $f_{\rm rh}$ could also be inferred by breaking this degeneracy, e.g. by additional theory motivated priors for $f_{\rm rh}$ or by requiring that the power spectrum does not overproduce PBH, to future studies.  
\begin{figure}
    \begin{center}
            \includegraphics[width=0.85\linewidth]{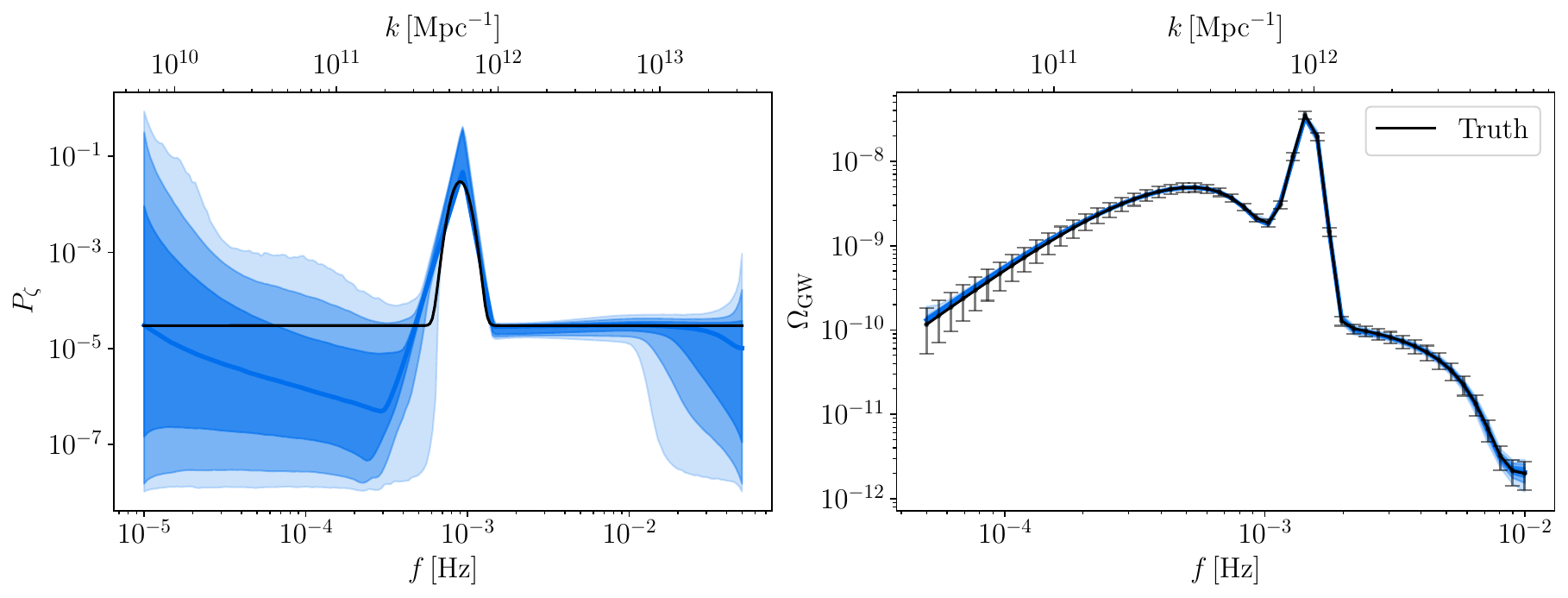}
    \end{center}
    \hspace{0.295\linewidth}%
\includegraphics[width=0.45\linewidth]{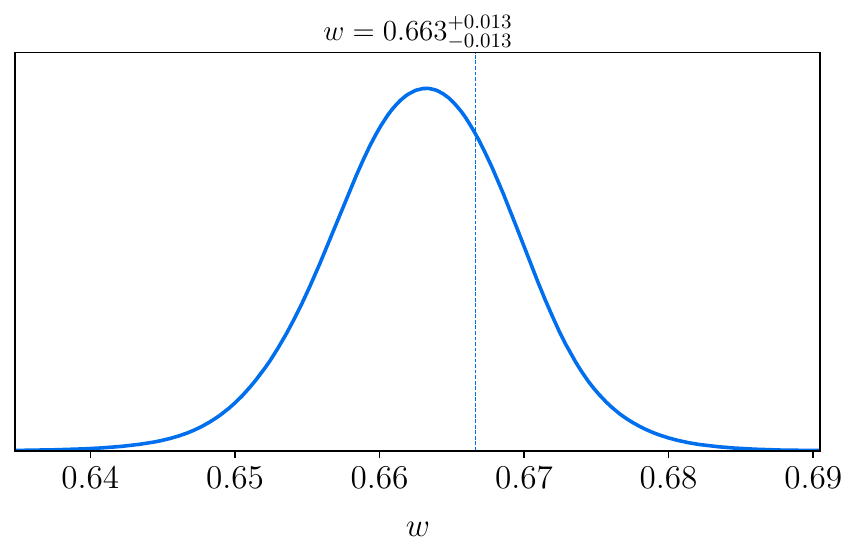} 
    \begin{center}
            \includegraphics[width=0.45\linewidth]{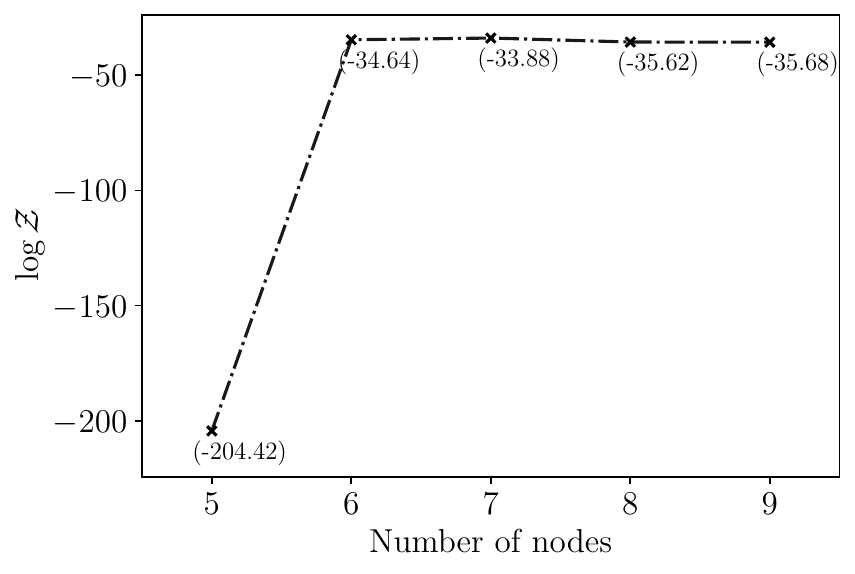}
    \end{center}
    \caption{\small Reconstruction of the peaked spectrum using linear interpolation for SIGW generated during an epoch with $w=2/3$, see section \ref{sec:gen_w}. The different shaded regions correspond to $68\%, 95\%$ and $99.7\%$ credible intervals for $\mP_\zeta$ and $\Omega_{\GW}$. Reported limits represent to $95\%$ credible intervals for the parameter $w$.}
    \label{fig:peaked_w}
\end{figure}
We present  the results for our fiducial model, which has the peaked and BPL templates  of sections \ref{sec_shapeak} and 
\ref{sec:bpl_res}
as  underlying curvature power spectra, while we choose $w=2/3$
as fiducial value for the equation of state. We use the same priors on the node amplitude and locations as in \cref{sec:RD} while for the equation of state we take $w\in [0,1]$. 

In \cref{fig:peaked_w}
we  plot the results for the case of peaked spectrum. We notice that  the underlying curvature power spectrum as well as the equation of state can be recovered quite well. The reason 
of such good reconstruction  is that for sharply peaked spectra, the position of the resonance peaks and dips in the SIGW spectra is strongly sensitive to the equation of state $w$. Thus, $w$ is recovered not only without any significant bias but also with tiny uncertainties. 

\begin{figure}
    \begin{center}
            \includegraphics[width=0.85\linewidth]{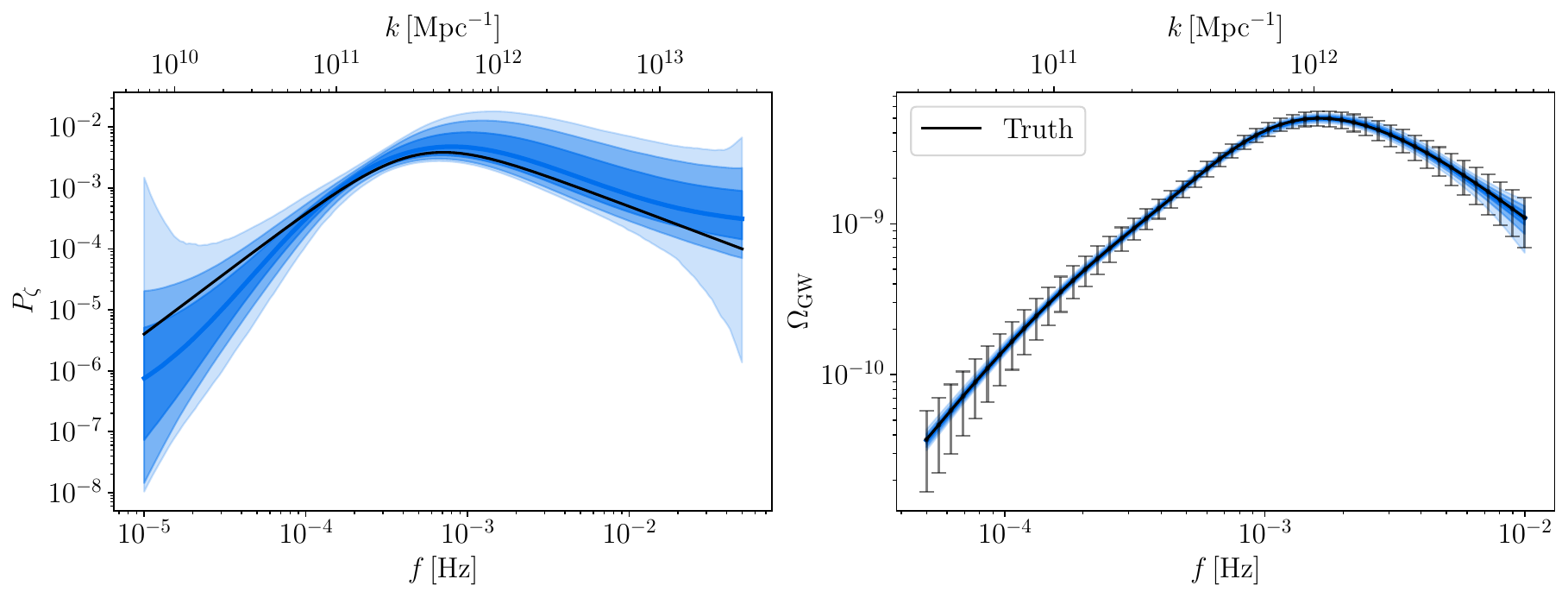}
    \end{center}
    \hspace{0.295\linewidth}%
    \includegraphics[width=0.45\linewidth]{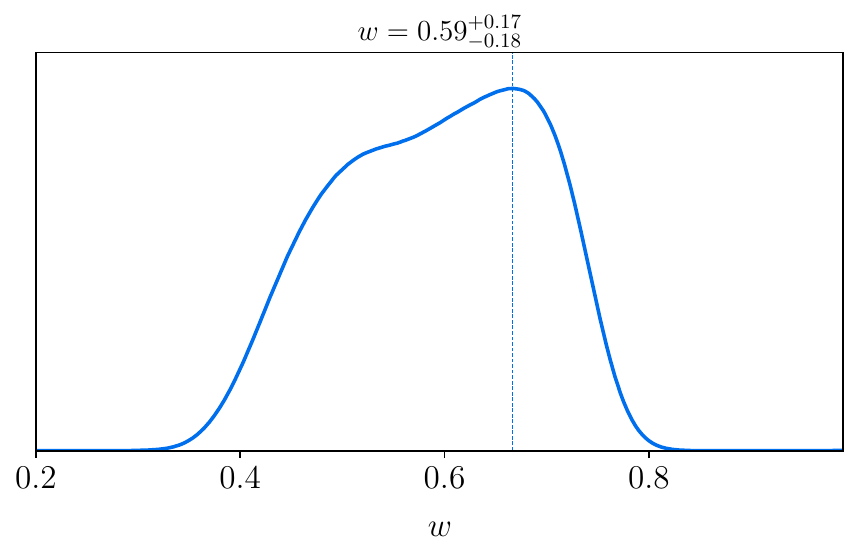} 
    \begin{center}
            \includegraphics[width=0.45\linewidth]{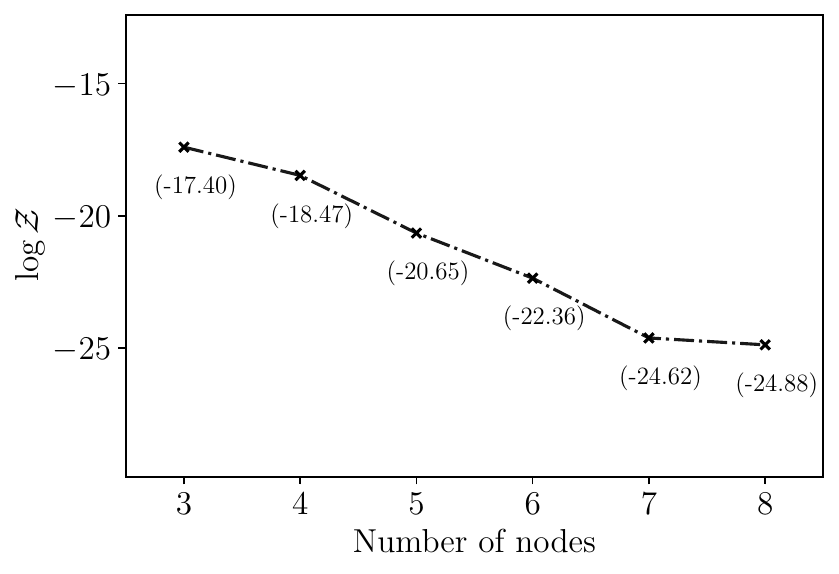}
    \end{center}
    \caption{\small Reconstruction of the BPL spectrum using GP interpolation for SIGW generated during an epoch with $w=2/3$, see section \ref{sec:gen_w}. The different shaded regions correspond to $68\%, 95\%$ and $99.7\%$ credible intervals for $\mP_\zeta$ and $\Omega_{\GW}$. Reported limits represent to $95\%$ credible intervals for the parameter $w$.}
    \label{fig:bpl_w}
\end{figure}

In \cref{fig:bpl_w}, we plot our results for our BPL model. Contrary to the previous examples, for this particular example we  use Gaussian Process (GP) interpolation to represent the power spectrum.\footnote{The reason being  that, for  the case of broad peaked spectra, linear interpolation  suffers from a degeneracy between the shape of $\mP_\zeta$ and $w$, owing to its greater flexibility, which can result in a bias in the inferred $w$ and $\mP_\zeta$.  The choice of GP interpolation restricts this degeneracy, by generating only smoothly varying $\mP_{\zeta}$. We expect that if one were to also directly reconstruct the inflationary potential itself instead of $\mP_{\zeta}$, such degeneracies would not arise since the resulting $\mP_{\zeta}$ would naturally be smooth. We discuss in more detail this subject  in \cref{app:broad_w}.}

In addition to the positions of the nodes and their amplitudes, GP interpolation  requires the specification of a  kernel function $K$. 
We adopt a radial basis function as the choice for the kernel:
\begin{align}
    K(x,x') = \exp\left[-\frac{(x-x')^2}{2l^2}\right]\,,
\end{align}
where \( x \) and \( x' \) denote the positions of the nodes\footnote{Not to be confused with $x=k\eta$ in section \cref{sec:sigwkernels}.}, once again in   \(\log_{10} k\)-space and 
$l$ corresponds to the GP lengthscale, which controls the correlation length of the GP predictions. Given positions of nodes $X$ and amplitudes $Y$, the prediction of the GP interpolation at a point $x_*$ is~\cite{williams2006gaussian}
\begin{align}
    f(x_*|X,Y) = K(x_{*},X)\cdot\,\bigl[K(X,X) + \sigma_n^2 I\bigr]^{-1}\cdot\,Y\,.
\end{align}
The term $\sigma^2_n I$ is a small number, added to the diagonal of the GP kernel matrix $K(X,X)$ 
to ensure numerical stability. The GP prediction is
a weighted mean of the values at the node locations, with the weights controlled by the GP lengthscale --  longer lengthscales means that nodes far from the point $x_*$ also contribute significantly to the prediction, and vice versa. We keep the same priors on the node locations and amplitudes as in the linear interpolation examples.  For the lengthscale, the prior reads $l \in [\Delta \log_{10}k /(8\times n), \Delta \log_{10}k]$ where $\Delta \log_{10}k$ is the width of the interval (in $\log_{10}k$ space) over which $\mP_{\zeta}$ is reconstructed and $n$ is the number of nodes. This lengthscale prior ensures smoothness of the interpolation while not being overly restrictive.
Following this procedure, we obtain that both the primordial power spectrum as well as the equation of state are recovered fairly well -- see \cref{fig:bpl_w}. However, we have  larger uncertainties in the inferred $w$ as opposed to the sharply peaked case of \cref{fig:peaked_w}. This is due to the fact that due to the broad shape of the peak in $\mP_\zeta$, the resonant peaks in $\Omega_{\GW}$ are absent.  The previous discussion of \cref{sec:bpl_res}  with respect to the behaviour of the IR and UV tails  also applies here,  with  only slight differences: \cref{eq:approx_bpl_RD} now changes to~\cite{Domenech:2021ztg}
\begin{align}
\!\!\!\!\!\Omega_{\mathrm{GW}}(k)\;\propto\;
\begin{cases}
\displaystyle
\biggl(\frac{k}{k_{\rm p}}\biggr)^{3}, & \!k \ll k_{\rm rh},\\[1ex]
\displaystyle
\biggl(\frac{k}{k_{\rm p}}\biggr)^{3 -2|b|}, &  \!k_{\rm rh} \ll k \ll k_{\rm p},\\[1ex]
\displaystyle
\biggl(\frac{k}{k_{\rm p}}\biggr)^{-\Delta -2|b|}, &\! k \gg k_{\rm p},
\end{cases}
\quad
\text{with}
\quad
\Delta =
\begin{cases}
2\,n_{\mathrm{UV}}, & \!\!\!0 < n_{\mathrm{UV}} +b < 4,\\[0.5ex]
4 + n_{\mathrm{UV}},   & \!n_{\mathrm{UV}} + b > 4,
\end{cases}
\label{eq:approx_bpl_w}
\end{align}
where $k_{\rm p}$ is the peak scale.

\section{Conclusions}
\label{sec:conclusions}

Scalar-induced gravitational waves (SIGW)
 produced in the early universe carry the imprints of the underlying curvature power spectrum as well as the cosmological background dynamics at the time of GW production. Since the production of
 SIGW is a process occurring at second 
order in fluctuations, it is not easy to infer 
properties of the source from measurements of the GW spectrum.  
In this work we have explored a Bayesian approach to reconstruct the power spectrum of the primordial curvature perturbation as well the equation of state from observations of SIGW. Our method is based on using interpolating splines to represent the  scalar power spectrum, with the number of nodes of the interpolation as well as their positions and amplitudes as parameters to be inferred from the data. 

We applied our method to specific mock $\Omega_{\GW}$ data, generated to be representative of the different kinds of shapes of the spectrum that can arise in well-motived early-universe models.
We showed that our method  is able to accurately reconstruct $\mP_{\zeta}$, especially if the peak of the SIGW is observed.
However, in cases where only the infrared part of the SIGW spectrum is observed, the reconstruction suffers due to the infrared universal scaling of the  spectrum of $\Omega_{\GW}$ for broad classes of $\mP_{\zeta}$ shapes, leading to mostly prior dominated constraints. 
As a specific example, we applied our techniques to reconstruct
the scalar spectrum leading to a GW signal fitting recent
PTA data. 
We also tested the viability of our approach in inferring the background equation of state (assuming a perfect fluid) at the time of SIGW production, alongside the power spectrum reconstruction. The shape of the SIGW spectrum depends strongly on $w$, but there also exist complex degeneracies between the effects of varying $w$ and varying $\mP_{\zeta}$, which we discussed. 

\medskip

Our method can be extended and applied to more general situations. It would be interesting to include
the effects of primordial curvature non-Gaussianities 
which can influence the SIGW profile (see e.g.~\cite{Unal:2018yaa,Cai:2018dig,Adshead:2021hnm,Perna:2024ehx}).
Besides curvature fluctuations, also isocurvature
modes might play a role in SIGW generation (see e.g.~
\cite{Domenech:2023jve,Passaglia:2021jla,Domenech:2021and,Marriott-Best:2025sez,Ozsoy:2023gnl,Chen:2024twp}). Although we focussed
on scalar sources, primordial tensor degrees of freedom can induce GW as well \cite{Bari:2023rcw}. Besides
considering more general primordial sources, it would also
be interesting to model more accurately our priors
regarding the infrared parts of the curvature
and GW spectra; to include accurate constraints related
to the corresponding primordial black hole formation;
and to analyse how imperfect knowledge of instrumental noise  might affect our results. Finally, suitable extensions of our method can allow not only to infer the $\mP_\zeta$ profile, but also to reconstruct the inflationary scalar potential
leading to SIGW profiles. We hope to return soon to these
topics in separate publications.

\section*{Acknowledgements}
We thank Sukannya Bhattacharya, Jonas El Gammal, Gabriele Franciolini, Juan Garcia-Bellido, Jan Hamann,  Marco Peloso, and David Wands
 for useful discussions. We are partially funded by the STFC grants ST/T000813/1 and ST/X000648/1. A.G. is supported by the UKRI AIMLAC CDT, funded by grant EP/S023992/1. We also acknowledge the support of the Supercomputing Wales project, which is part-funded by the European Regional Development Fund (ERDF) via Welsh Government. 
For the purpose of open access, the authors have applied a Creative Commons Attribution licence to any Author Accepted Manuscript version arising. Research Data Access Statement: The code used to generate the mock data and run the analysis is available at \cite{Ameek94SIGWInverse}.

\begin{appendix}

\section{Scalar power spectrum reconstruction with recent PTA results}
\label{app_pta}

\begin{figure}[ht]
    \centering
    \includegraphics[width=0.85\linewidth]{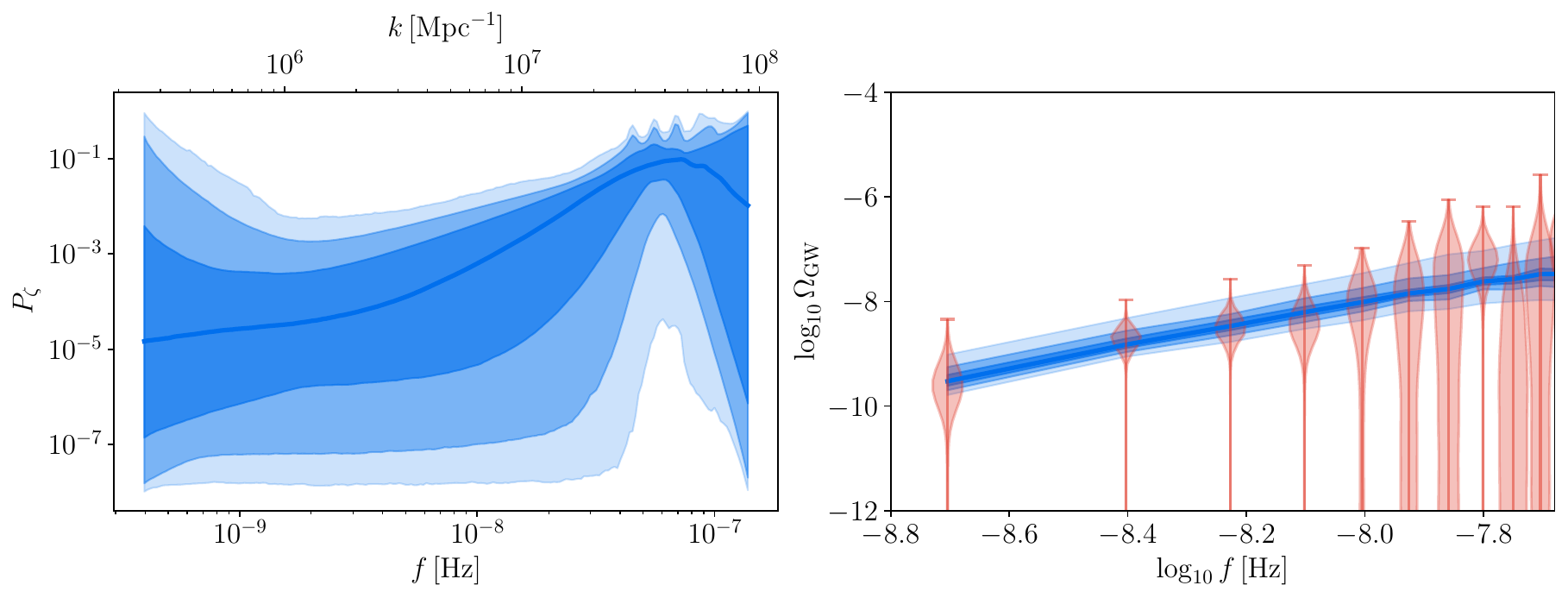}
    \caption{\small Spline based reconstruction of the curvature power spectrum from PTA (NANOGrav 15 year). \textit{Left}: The power spectrum posterior \textit{Right}: The corresponding $\Omega_{\rm GW}$ spectrum along with violin plots of the free spectrum Hellings Downs reconstruction from NG15.}
    \label{fig:NG15}
\end{figure}

We  test our method against real data from Pulsar Timing Arrays.\footnote{
See e.g.~\cite{NANOGrav:2023hvm,EPTA:2023xxk,Figueroa:2023zhu,Ellis:2023oxs,Cai:2023dls,Yi:2023mbm,Yi:2023tdk,Firouzjahi:2023lzg,Liu:2023pau,Wang:2023ost,Domenech:2024rks,Harigaya:2023pmw} for other studies.
We note that the scalar induced GW interpretation of the PTA results is likely ruled out due to overproduction of PBH, at least in the case of Gaussian statistics of $\zeta$ and assuming radiation domination. The inclusion of non-Gaussianity or a non-standard equation of state may alleviate this, as studied in refs~\cite{Balaji:2023ehk,Wang:2023ost,Franciolini:2023pbf,Harigaya:2023pmw,Domenech:2024rks,Liu:2023ymk,Liu:2023hpw,Inui:2024fgk}.} We make use of the public package \texttt{Ceffyl}~\cite{lamb2023rapid} to interface our code with the NANOGrav 15 year data~\cite{NANOGrav:2023gor,the_nanograv_collaboration_2025_16051178}. The results of the reconstruction are plotted in \cref{fig:NG15}. We do not find any evidence for significant deviations from a power-law $\mP_{\zeta}$ or $\Omega_{\GW}$ with little to choose amongst the different models, with the $\log \zc = -63.71, -63.66-63.60, -63.59, -64.57$ for $N = 2,3,4,5,6$ respectively.

\section{General equation of state and broad spectra}
\label{app:broad_w}
\begin{figure}[h]
    \centering
    \includegraphics[width=0.85\linewidth]{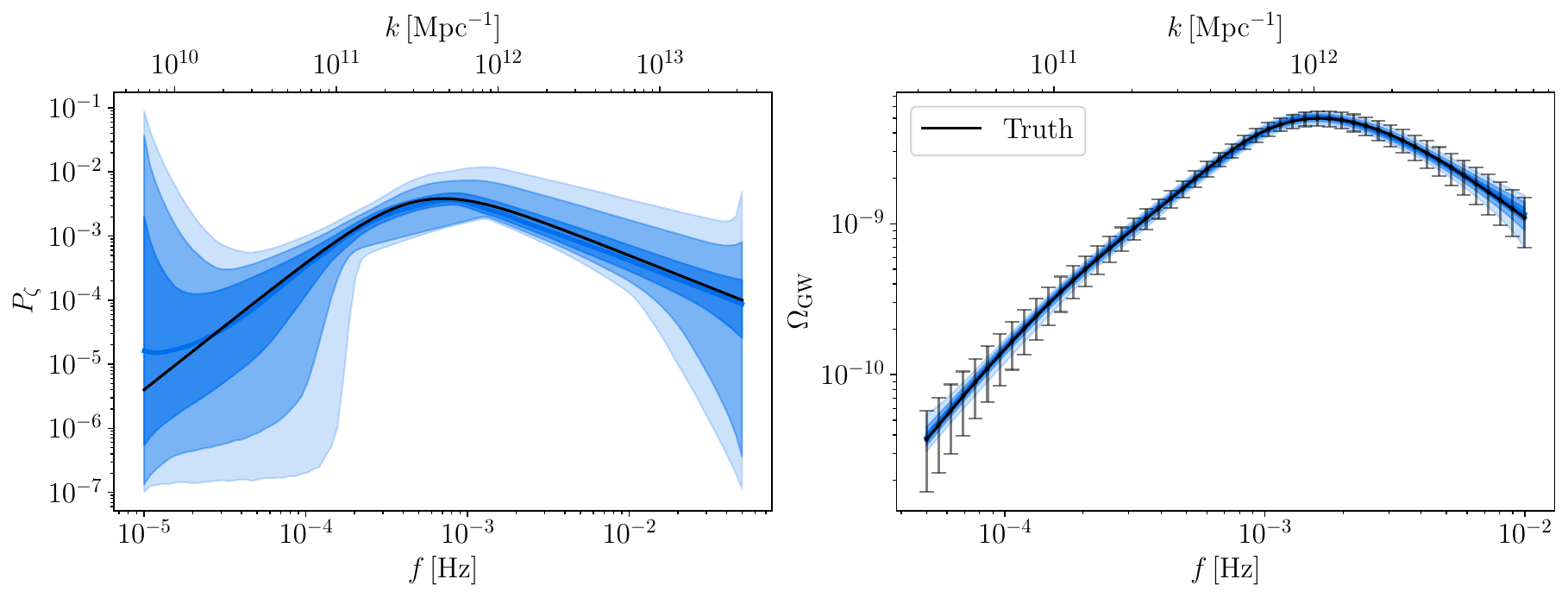}
    \includegraphics[width=0.45\linewidth]{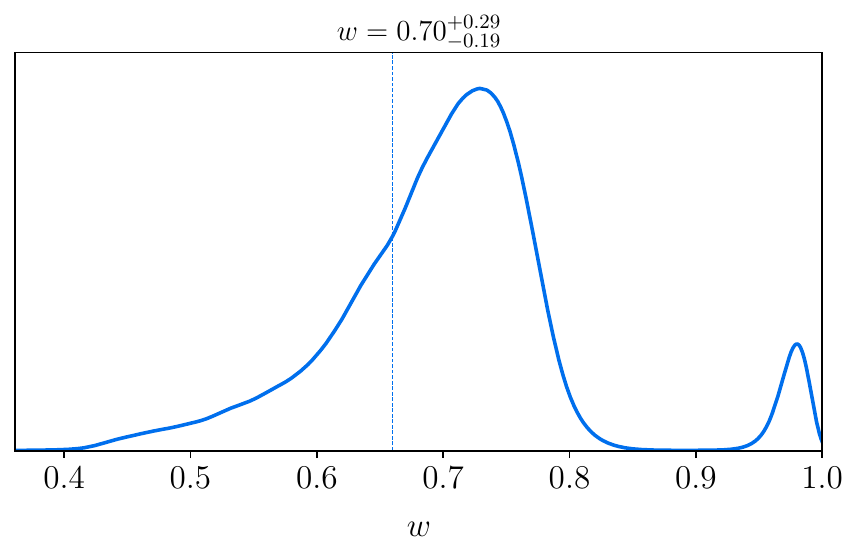}
    \caption{\small Reconstruction of the BPL example using linear interpolation with five nodes, as  
    discussed in appendix \ref{app:broad_w}.}
    \label{fig:bpl_w_double_peak}
\end{figure}

In this section we present results for the BPL example with $w=2/3$, reconstructed using linear interpolation. We see that in addition to a peak in the posterior distribution of $w$ at the correct value, there is an additional peak towards the $w=1$. The reason behind this is that as $w$ increases, the position of the peak in the SIGW spectrum shifts towards higher frequencies, and as $w,c_s\to 1$ the peak also starts to broaden. These effects are then compensated by $\mP_\zeta$ shapes with sharp variations across the $k$-range of the reconstruction (visible around $f\sim 10^{-4}$ Hz in the left panel of \cref{fig:bpl_w_double_peak}), which bear very little resemblance to the true curvature power spectrum.  This is a consequence of the additional flexibility of the linear interpolation, as opposed to the smooth Gaussian process interpolation we employed for this example in \cref{sec:gen_w}. We expect direct reconstruction of the inflationary potential to also alleviate this issue, since in that case as well the resulting power spectrum would be much smoother compared to the linear interpolation one.

One might wonder why there is such a bias in the inferred value of $w$, given the dependence of the shape of the SIGW spectrum on $w$ as seen from \cref{eq:approx_bpl_w}. In particular, for the BPL shape of the primordial curvature power spectrum, the IR part of the SIGW spectrum is indpendent of the IR behaviour of $\mP_{\zeta}$ and is determined entirely by $w$. However, the reason behind the bias is that for this example, the chosen error bars on $\Omega_{\GW}$ imply that the IR slope of the SIGW spectrum is not precisely determined, which allows for different values of $w$ other than the true one, to be compatible with the data.

To test this, we generate $\Omega_{\GW}$ data with much lower noise values, particularly in the tails. This is given by
\begin{align}
\label{eq:DOGW_low}
    \Delta \Omega_{\rm GW}(f) = \Omega_{\rm GW ,obs}(f) \left[0.05 + 0.01(\log f/f_*)^2  \right]\,.
\end{align}
This is to be contrasted with \cref{eq:DOGW}, which represented the magnitude of the error bars chosen for all other examples (except for PTA) in this work. 
\begin{figure}[h]
    \centering
    \includegraphics[width=0.45\linewidth]{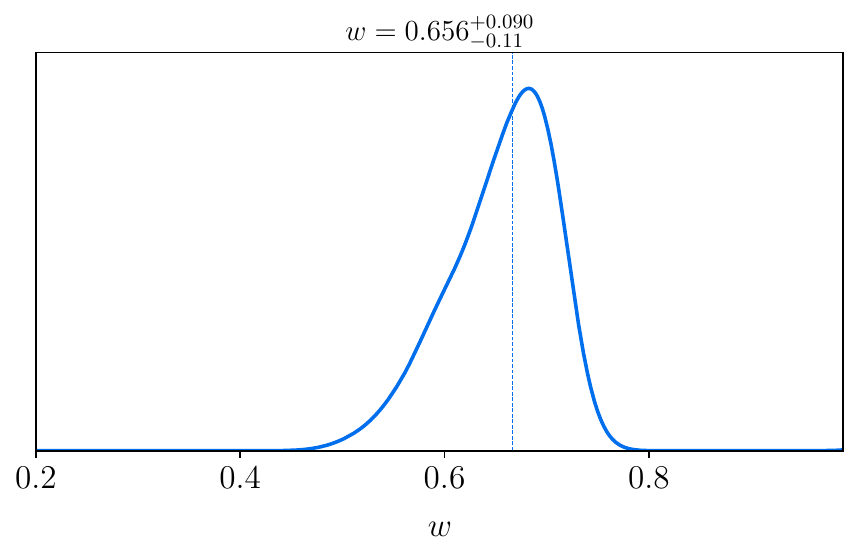}
    \caption{Reconstruction using linear interpolation for the low noise BPL example with $w=2/3$. }
    \label{fig:bpl_w_low_noise}
\end{figure}
We show in \cref{fig:bpl_w_low_noise} the result of the $w$ reconstruction for this low noise example where we see that $w$ is now inferred correctly without any bias due to the IR slope of the SIGW spectrum being measured much more precisely.\footnote{The degeneracies persist for the oscillatory model of \cref{eq:osc} and are in fact much more severe, leading to bias in the inferred $w$ even with the lower noise level example considered here.}

\section{Noisy realisations of GW data}
\label{app:noisy_realisations}
{
To test the robustness of our method, we also apply it to noisy realisations of the mean spectra presented in \cref{sec:RD} for which we generate random realisations drawn from a Gaussian distribution defined by \cref{eq:Gaussian_likelihood}. Our method can again recover the a curvature power spectrum close to the actual one, as long as the noise level close to the peak of the GW spectrum is low i.e. the shape of the peak of the noisy realisation is not too different from the mean spectrum. As discussed earlier, this is a consequence of the fact that the shape of the induced GW spectrum depends most strongly on the shape of the peak of the curvature power spectrum. The latter itself gets tightly constrained by the low noise levels around the peak of the GW spectrum. 

The Bayesian evidence shows a clear trend for the BPL model, irrespective of the individual realisation, with models with more than 6 nodes being favoured less, as shown in Figure \ref{fig:BPL_MC_LogZ}. For brevity, we only show the reconstruction for one particular realisation. {We also plot the fidelity of the reconstructed $P_\zeta$ and $\Omega_{\rm GW}$ for this realisation, showing $\Delta P_\zeta \equiv |\hat{P}_\zeta - P_{\zeta}^{\rm true}|/P_{\zeta}^{\rm true}$ where $\hat{P}_{\zeta}$ represents the median of the reconstruction and similarly for $\Omega_{\rm GW}$. We see that the relative error remains small, reaching $10^{-1}$ or lower near the peak and only gets higher far away from it, especially in the IR for $P_\zeta$.}
\begin{figure}[h]
    \centering
\includegraphics[width=0.85\linewidth]{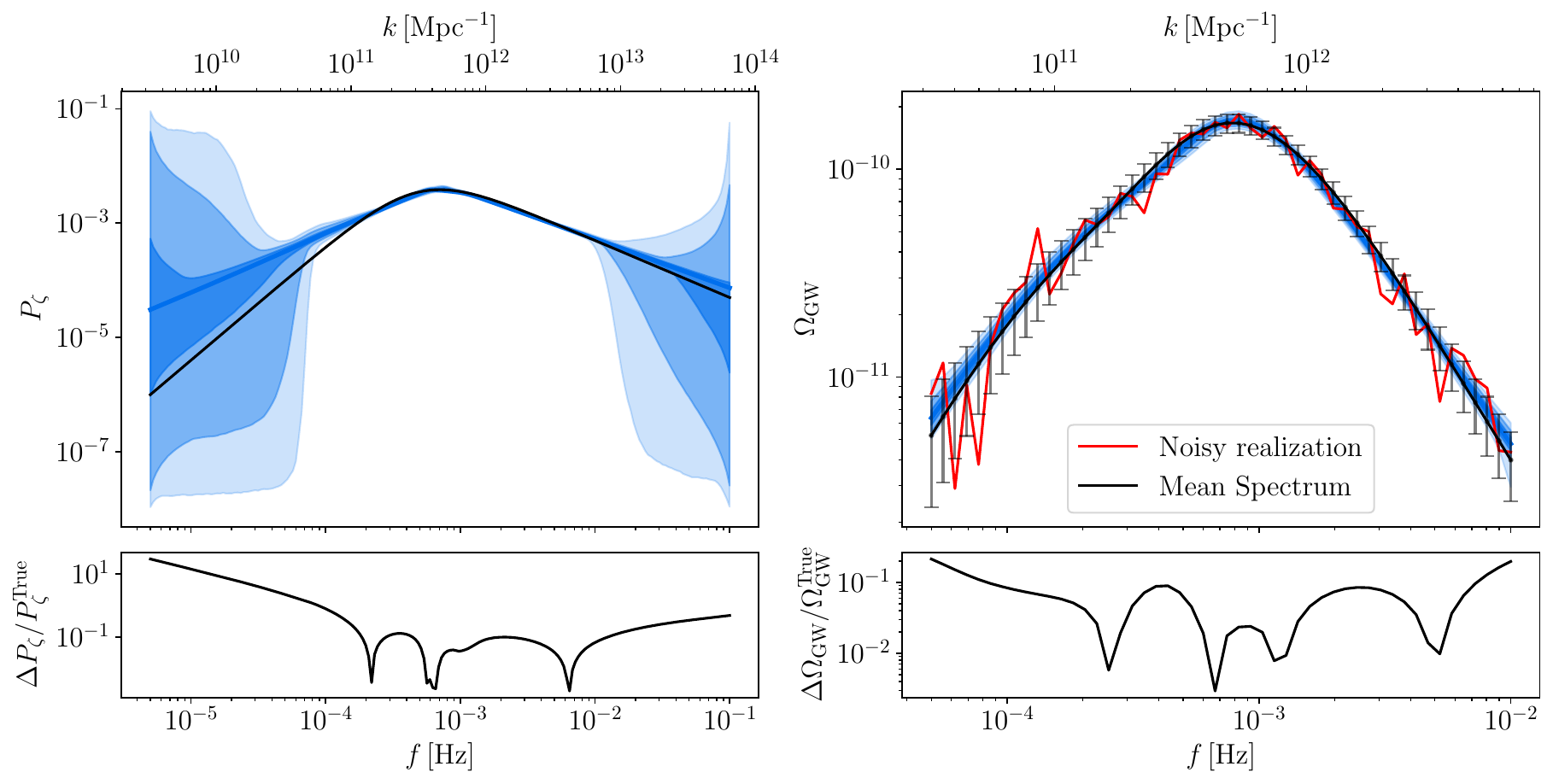}
    \includegraphics[width=0.4\linewidth]{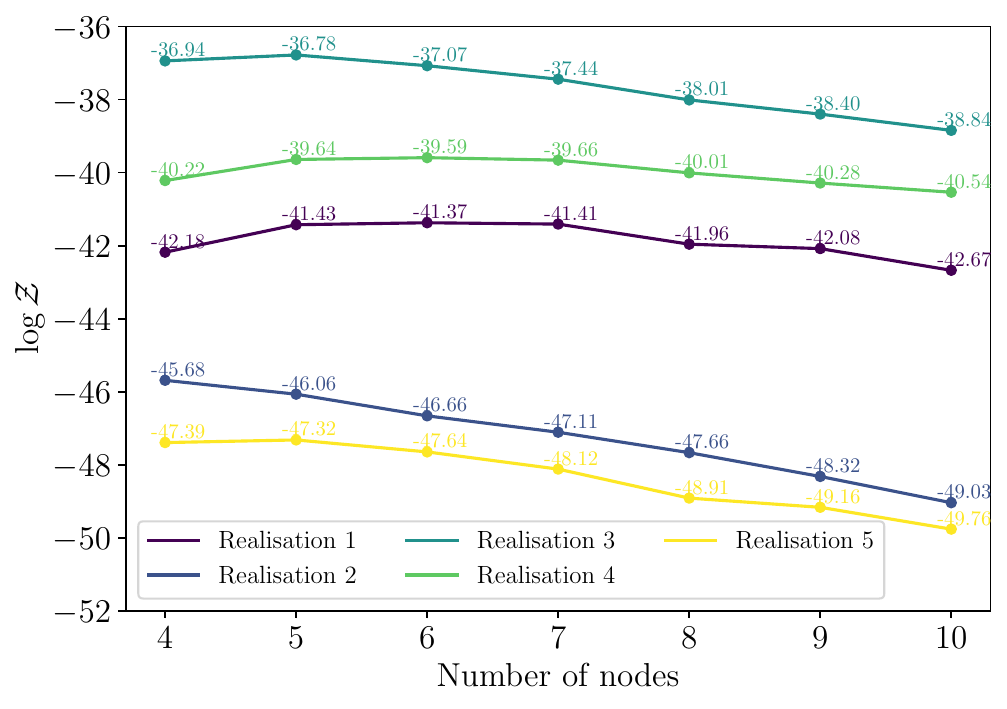}
    \caption{{\textit{Top}: Reconstruction of the BPL spectrum for a particular noisy realisation of the GW data using linear interpolation. \textit{Bottom}: $\log \zc$ values across five different realisations.}}
    \label{fig:BPL_MC_LogZ}
\end{figure}

For the peaked and oscillatory spectra we find that we are still able to accurately recover the underlying curvature power spectrum. However, increasing the number of nodes does not immediately lead to a decrease of the evidence independently of the random realisation, owing to the increased complexity of the models, similar to what we observed with the mean-only realisations in \cref{sec:RD}}.

\end{appendix}

\newpage
{\small
\addcontentsline{toc}{section}{References}
\bibliographystyle{utphys}
\bibliography{refs}
}

\end{document}